\documentclass[]{article}

\usepackage{amsmath}
\usepackage{amssymb}
\usepackage{tikz}
\usepackage{graphicx}
\usepackage[footnotesize,hang]{caption}
\usepackage{subfig}
\usepackage{color}
\usepackage{rotating}
\usepackage{mathtools}
\usepackage{xcolor}
\usepackage{pifont}

\usepackage[margin=0.85in]{geometry}

\usetikzlibrary{decorations.pathmorphing}

\renewcommand{\inf}{\infty}
\newcommand{\eps}{\epsilon}

\renewcommand{\i}{\mathrm{i}}
\newcommand{\e}{\mathrm{e}}

\renewcommand{\d}{\,\mathrm{d}}
\newcommand{\diff}[2]{\frac{\mathrm{d} #1}{\mathrm{d} #2}}
\newcommand{\sech}{\mathrm{sech}}

\providecommand{\keywords}[1]{\textbf{\textit{Keywords:}} #1}
\providecommand{\AMS}[1]{\textbf{\textit{AMS Subject Classifications:}} #1}

\def\XXint#1#2#3{{\setbox0=\hbox{$#1{#2#3}{\int}$}
\vcenter{\hbox{$#2#3$}}\kern-.5\wd0}}

\title{Nanoptera and Stokes Curves in the 2-Periodic Fermi-Pasta-Ulam-Tsingou Equation}
\author{C. J. Lustri$^1$\footnote{Electronic address: christopher.lustri@mq.edu.au}}
\date{%
    $^1$Department of Mathematics and Statistics, 12 Wally's Walk, Macquarie University, New South Wales 2109, Australia\\[2ex]%
}                                     % Activate to display a given date or no date

\begin{document}
\maketitle

\begin{abstract}
This work presents asymptotic solutions to a singularly-perturbed, period-2 FPUT lattice and uses exponential asymptotics to examine `nanoptera',  which are nonlocal solitary waves with constant-amplitude, exponentially small wave trains which appear behind the wave front. Using an exponential asymptotic approach, this work isolates the exponentially small oscillations, and demonstrates that they appear as special curves in the analytically-continued solution, known as `Stokes curves' are crossed. \textcolor{black}{By studying the asymptotic form of these isolated oscillations}, it is shown that there are special mass ratios which cause the oscillations to vanish, producing localized solitary-wave solutions. The asymptotic predictions are validated through comparison with numerical simulations.
\end{abstract}

%%%%%

%\begin{keywords}
%\end{keywords}

%%%%%

%\begin{AMS}
\AMS{34E15, 35Q51, 34C15, 37K60}
%\end{AMS}

\section{Introduction}\label{S.Intro}

In this study, we demonstrate the important role played by Stokes phenomenon in the behaviour of travelling waves propagating through a diatomic Fermi-Pasta-Ulam-Tsingou (FPUT) lattice. The classical FPUT lattice contains an infinitely long line of masses connected to their neighbours by identical springs. After non-dimensionalization, this system can be represented by
\begin{align}\label{1.FPUT}
m(j) \ddot{x}(j,t) = F(x(j+1,t) - x(j,t)) - F(x(j,t) - x(j-1,t)),
\end{align}
where $x(j,t)$ represents the position of the $j$th particle at time $t$, a dot refers to differentiation with respect to time, and $F$ is an interaction \textcolor{black}{force} given by
\begin{equation}\label{1.POT}
F(r) = r + r^2.
\end{equation}
A diatomic lattice has $m(j) = m_1$ if $j$ is odd, and $m(j) = m_2$ if $j$ is even, where $m_1 \neq m_2$. This system has a speed of sound given by $c_0 = \sqrt{2/(m_1+m_2)}$. In this study, we are interested in the behaviour of supersonic travelling waves (with speed $c_{\epsilon} > c_0$) in diatomic systems with small mass ratio. We therefore set $\delta$ such that $m_2/m_1 = \delta^2$ with $0 < \delta \ll 1$. 

The behaviour of travelling waves in monoatomic FPUT lattice has been treated comprehensively in, for example, \cite{Friesecke1,Friesecke2,Friesecke3}. These studies determined that the travelling wave solution is a localized solitary wave that is a regular perturbation away from a solution to the Kortweg-de Vries (KdV) equation in the long-wave asymptotic limit. These studies led to the natural question of whether the results could be generalized to polyatomic FPUT lattices, and how travelling waves would propagate through such systems.

% Note the speed of sound from Freischke and Watts/Pego (see Hoffman and Wright). We are considering small fixed mass, with a speed just above the speed of sound, as in Faver \& Wright.

In a later analysis, Gaison \textit{et al.} \cite{Gaison1} found that the equations governing travelling waves in diatomic FPUT systems can be written as a singular, rather than regular, perturbation to a solution to the KdV equation. The KdV solution is a long-wave approximation that depends on a small parameter, and the solution is accurate up to algebraic order in this parameter. We will later use this KdV solution as a leading-order solution to the diatomic FPUT lattice equation \eqref{1.FPUT} in the small mass ratio limit. A number of other rigorous and formal studies of the behaviour of periodic lattice systems have been undertaken in the past, including \cite{Chirilus1,Pnevmatikos1,Porter1,Schneider1,Yong1}.

\begin{figure}
\centering
\subfloat[Solitary Wave]{
\begin{tikzpicture}
[xscale=0.4,>=stealth,yscale=-2]
\draw[gray] (-5,0.5) node[left] {\scriptsize{$0$}} -- (5,0.5);
\draw[black,thick] plot[smooth] file {Sol1b.txt}; 
\draw (-5,1.2) -- (5,1.2) -- (5,-0.2) -- (-5,-0.2) -- cycle;
\draw [->] (1,0.25) -- (2,0.25) node[ right] {\scriptsize{$x = ct$}};
\end{tikzpicture}
}
\subfloat[One-sided nanopteron]{
\begin{tikzpicture}
[xscale=0.4,>=stealth,yscale=-2]
\draw[gray] (-5,0.5) -- (5,0.5);
\draw[black,thick] plot[smooth] file {Sol2b.txt}; 
\draw (-5,1.2) -- (5,1.2) -- (5,-0.2) -- (-5,-0.2) -- cycle;
\draw [->] (1,0.25) -- (2,0.25) node[ right] {\scriptsize{$x = ct$}};
\end{tikzpicture}
}
\subfloat[Two-sided nanopteron]{
\begin{tikzpicture}
[xscale=0.4,>=stealth,yscale=-2]
\draw[gray] (-5,0.5) -- (5,0.5);
\draw[black,thick] plot[smooth] file {Sol3b.txt}; 
\draw (-5,1.2) -- (5,1.2) -- (5,-0.2) -- (-5,-0.2) -- cycle;
\draw [->] (1,0.25) -- (2,0.25) node[ right] {\scriptsize{$x =ct$}};
\end{tikzpicture}
}
\caption{Comparison of the profiles associated with (a) a standard solitary wave, (b) a one-sided nanopteron, and (c) a two-sided nanopteron that each propagate at speed $c$. The solitary wave is localized spatially, whereas the nanoptera have non-decaying oscillatory tails on (b) one side or (c) both sides of the wave front. \textcolor{black}{Only the waves shown in (a) and (c) are truly steady, while the profile shown in (b) is metastable; the leading-order travelling wave profile and oscillations are steady, but the one-sided radiation must draw energy out of the wave front, leading to its eventual decay over a long timescale that is not captured by the leading order asymptotics. }}\label{F:nanoptera}
\end{figure}
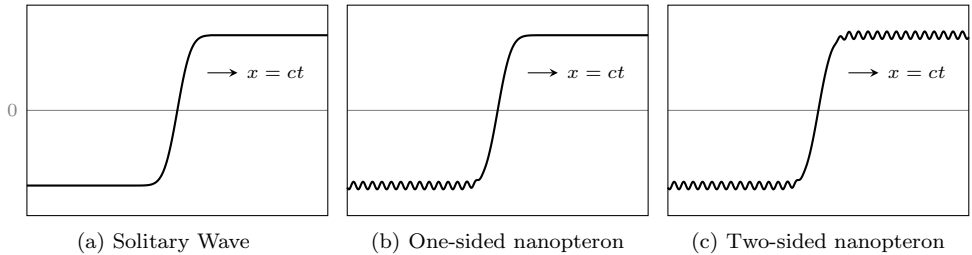

Faver \& Wright \cite{Faver2} and Hoffman \& Wright \cite{Hoffman1} used a Beale ansatz (introduced in \cite{Beale1}) to rigorously prove the existence of supersonic travelling wave solutions in the diatomic FPUT lattice, which are composed of the sum of an exponentially localized solitary wave and periodic oscillations with exponentially small amplitude. Solutions of this form are known as `nanopteron' solutions. Boyd introduced the term `nanopteron' in \cite{Boyd4} to describe weakly nonlocal solitary waves, which approximately satisfy the classical definition of a solitary wave with the distinction that a nanopteron wave is not exponentially localized, but rather tends to a small-amplitude oscillation on either one or both sides of the leading-order solitary wave, illustrated in Figure \ref{F:nanoptera}. Iooss \& Kirchg\"assner \cite{Iooss1} used related methods to identify nanopteron solutions in a chain of nonlinear oscillators with FPUT interactions and a potential function.

\textcolor{black}{Typical nanopteron solutions contain a leading-order travelling wave, and a train of exponentially small oscillations on one or both sides of this front. When these solutions were first identified, it was commonly assumed that waves of this form were typically steady, and that these waves could propagate indefinitely in the same manner as classical solitary waves. This idea was contradicted by the work of Boyd \cite{Boyd6}, who studied a fifth-order Korteweg-de Vries (KdV) and concluded that only symmetric two-sided nanoptera could truly propagate in this manner. One-sided nanoptera, or two-sided nanoptera with unequal amplitude on either side, must necessarily radiate energy and can therefore not propagate indefinitely. This argument was made explicitly for the fifth-order KdV in \cite{Benilov1}, and the analysis therein can equally be applied to diatomic lattice systems. }

\textcolor{black}{This is consistent with numerical studies of one-sided nanopteron solutions in \cite{Giardetti1,Jayaprakash1,Vainchtein1}, which showed numerical examples of one-sided nanoptera in diatomic FPUT, Hertzian, and Toda chains respectively, each of which radiated energy and eventually decayed. Notably, metastable nanopteron solutions to the diatomic FPUT lattice equation were directly computed in the numerical study \cite{Giardetti1}. This study presented the numerical technique used in Section \ref{S.Numerics} to validate the formal asymptotic analysis. }

\textcolor{black}{One-sided nanoptera are therefore not truly steady in these particle systems, but instead decay on a slow timescale in a fashion that is not visible in the leading-order approximation to the wavefront or the exponentially small oscillations. Giardetti \textit{et al.} \cite{Giardetti1} conjecture that this decay occurs in the diatomic FPUT system on a timescale that is exponentially large in the small mass ratio parameter. Hence, these solutions are not true travelling wave solutions, and are instead described as ``metastable'' or ``quasi-stable'' solutions.} 

\textcolor{black}{The detailed analysis in this paper will compute the behaviour of one-sided metastable solutions; however, it is straightforward to adapt the analysis to obtain a form for symmetric two-sided solutions, and this will also be noted at appropriate points in the analysis. The term ``nanopteron'' will be used to describe both stable waves with symmetric oscillations, as well as metastable solutions with waves on one side. In the latter case, the leading-order asymptotics provide a useful approximation to the wave behaviour in the quasi-steady regime.}

A number of other studies have been performed on diatomic FPUT (or related) lattices. Qin \cite{Qin1} proved the existence of periodic supersonic wavetrains in such a lattice, which correspond to the far-field periodic oscillations far from the travelling wave front in the present study. Faver later used a Beale ansatz to prove the existence of nanopteron solutions in FPUT lattices with 2-periodic spring potential \eqref{1.POT}, rather than the 2-periodic mass considered here.

Previous mathematical and experimental studies of lattices with periodic masses have revealed dynamics which do not arise in uniform lattices \cite{Jayaprakash1,Jayaprakash2,Okada1,Porter1,Tabata1,Vainchtein1}. In particular, diatomic lattices permit new families of solitary waves which can exist only for discrete values of the ratio between the masses of the two particles in a diatomic unit \cite{Jayaprakash1,Jayaprakash2,Vainchtein1}. In \cite{Jayaprakash1}, these ratios are described as anti-resonances. Asymptotic descriptions of these mass ratios for the diatomic Toda lattice were first obtained in \cite{Vainchtein1} using a matched asymptotic expansion technique, and explained in \cite{Lustri5} as destructive interference between two distinct exponentially small wavetrains in the solution. This study exploited the asymptotic form of the exponentially small terms to find a simplified asymptotic expression for the mass ratios that produce cancellation.

\textcolor{black}{More generally, exponential asymptotic methods have been used to discuss the behaviour of a wide range of systems of with discrete dynamics; this includes both differential-difference and difference equations. An early of exponential asymptotics being applied to a discrete problem is found in \cite{Tovbis1}, which uses Borel transform techniques to study exponentially small seperatrix splitting in the discrete H\'enon map. A number of subsequent studies used Borel transform methods to identify exponentially localized waves in discrete systems, including the study of radiationless travelling kinks in $\phi^4$ equations \cite{Oxtoby2} and sliding solitons in the discrete nonlinear Schr\"odinger (NLS) equation \cite{Oxtoby1}. Radiationless travelling waves were also identified in saturable NLS systems \cite{Melvin1}, FPUT oscillator chains \cite{Iooss1, Iooss2} and a nonlocal sine-Gordon model for Josephson structures \cite{Alfimov2}.  Recently \cite{Alfimov3} used similar exponential asymptotic techniques to the present study to compute exponentially small oscillations in soliton solutions to the discrete NLS equation, and compare these solutions to those of the continuous second-order NLS that obtained in the continuum limit; this analysis found that the localized solutions of the continuous system were not the same as the transparent points (or approximately localized solutions) of the discrete equation. Similar techniques were used in \cite{King4} to study dislocations in the discrete Frenkel-Kontorova model.}

\textcolor{black}{In these previous studies, exponential asymptotic techniques were used to identify localized solutions (otherwise known as ``sliding solutions'' or ``radiationless solutions'', as in \cite{Alfimov2, Oxtoby2,Oxtoby1}). Each system possesses travelling leading-order solutions that, in general, correspond to solutions with exponentially small oscillatory tails. By directly calculating the behaviour of these oscillations, the authors were able to identify special parameter sets in which they disappear entirely, producing a purely localized solitary wave. This can occur either because the amplitudes are proportional to a constant (known as ``Stokes constants'') which takes zero value for certain parameter choices, as in \cite{Melvin1,Oxtoby2,Oxtoby1}, or because there are multiple oscillatory contributions that interfere destructively for certain parameter choices, as in \cite{Alfimov3,Lustri5}. These wave-free solutions are true solitary waves which do not decay due to radiation being lost to the oscillatory tail.}

This paper uses an adaptation of the technique developed in \cite{Lustri5} based on \cite{Chapman1,Daalhuis1} for studying travelling waves in a diatomic Toda lattice with small mass ratio. This previous work found that travelling waves in the diatomic Toda are nanoptera, and that the exponentially small oscillations appear as curves in the complex plane known as `Stokes curves', discussed in Section \ref{S.AsympIntro}.  In the present study, we show that corresponding behaviour is found in supersonic travelling wave solutions for the diatomic FPUT lattice with small mass ratio. We will also show that the special nanopteron-free asymptotic solutions -- true solitary waves -- computed in \cite{Lustri5, Vainchtein1} are also present in the diatomic FPUT lattice, and are caused by destructive interference between two distinct oscillatory contributions.

In the present work, we do not obtain a rigorous existence proof, and we instead take a similar approach to \cite{Lustri5, Vainchtein1} and compare the results of our formal analysis to computational results. In this study, we use exponential asymptotic techniques developed by \cite{Chapman1, Daalhuis1} to provide a mathematical description of nanopteron solutions in the period-2 FPUT lattice. 

The remainder of this paper is organized as follows. The method of exponential asymptotics is introduced in Section \ref{S.AsympIntro}. The equations for diatomic FPUT lattices are shown in Section \ref{S.FPUT}, the long wave approximation to the solution is provided. In Section \ref{S.Series}, the solutions to the FPUT equations are expanded as asymptotic power series in the small mass ratio. The leading-order behaviour of this series is found in Section \ref{S.LOT}, and an ansatz is applied to determine the late-order behaviour in Section \ref{S.LateOrder}, with some technical details found in Appendix \ref{A.Lambda}. A full exponential asymptotic analysis is performed in Section \ref{S.ExpAsymp}. This involves finding the Stokes structure of the solution in \ref{S.StokesStruct}, and determining the asymptotic form of the exponentially small oscillations in Section \ref{S.Remainder}. In Section \ref{S.Numerics}, the asymptotic results are compared to numerical simulations, and it is noted that nanopteron-free solutions are apparent in both the numerical and asymptotic results. In Section \ref{S.Orthogonality}, an asymptotic expression is obtained for the special mass ratios that produce nanopteron-free solutions. The paper concludes in Section \ref{S.Conclusions}. 

\subsection{Exponential Asymptotics}\label{S.AsympIntro}

We examine the asymptotic behavior of exponentially small, non-decaying waves that appear in the wake of a solitary-wave front in diatomic FPUT lattices. However, determining the behavior of terms that are exponentially small compared to the leading-order solution in the $\eps \rightarrow 0$ asymptotic limit is impossible using classical asymptotic series expansions, because the exponentially small contribution is necessarily smaller than any power of the small parameter $\delta$. Therefore, we apply specialized techniques, known as `{exponential asymptotics}', to determine behavior on this scale \cite{Boyd3,Boyd1}. This section contains a brief outline of the method that will be used in the present study, adapted from the description in \cite{Lustri5}.

As we uncover this exponentially small behavior, we will see that the analytic continuation of the asymptotic solution includes curves known as `Stokes curves' \cite{Stokes1}. These curves are related to the behavior of exponentially small contributions to the solutions. As a Stokes curve is crossed, the exponentially small contribution experiences a smooth, rapid change in value in the neighborhood of the curve. In many problems, including the present investigation, the exponentially small contribution to the solution appears only on one side of a Stokes curve.

The central idea of exponential asymptotic methods is that a divergent asymptotic series may be truncated to approximate the exact solution. Furthermore, one can choose the truncation point to minimize the error between the approximation and the exact solution; this is known as `{optimal truncation};. When a divergent series is truncated optimally, the approximation error is generally exponentially small in the asymptotic limit \cite{Boyd1}. The problem may then be rescaled to directly determine this approximation error, allowing the exponentially small component of the solution to be determined in the absence of the asymptotic series itself. This idea was introduced by Berry \cite{Berry1,Berry4}, and it was employed in \cite{Berry3,Berry5} to determine the position of Stokes curves, and associated switching behaviour in special functions such as the Airy function. 

In the present paper, we apply an exponential asymptotic method developed by Olde Daalhuis et al. \cite{Daalhuis1} for linear differential equations and extended by Chapman et al. \cite{Chapman1} to nonlinear ordinary differential equations. Here we provide a brief outline of the process; see the above papers for a more detailed explanation of the methodology.

The first step in exponential asymptotic analysis is to express the solution as an asymptotic power series. In many singular perturbation problems, including the problem considered in the present study, the asymptotic series solution diverges. For a more detailed discussion of asymptotic series divergence, see \cite{Boyd2,Dingle1}. Optimally truncating these divergent series typically requires a general form for the asymptotic series coefficients, and it is frequently algebraically intractable to obtain such a general form. In practice, however, one does not require the exact form of the series coefficients. Instead, one needs only the so-called `{late-order terms}', or asymptotic expressions for the $r$th series coefficient in the $r \rightarrow \inf$ limit.

Dingle \cite{Dingle1} noted that the terms of the divergent asymptotic power series of a singularly perturbed system are typically obtained by repeated differentiations, and therefore diverge in a predictable factorial-over-power fashion. Noting this observation, Chapman et al. \cite{Chapman1} proposed writing an ansatz for the late-order terms that is capable of describing this form of late-order term behavior. One can write an ansatz for the $r$th term of a divergent asymptotic series (denoted $g_r$) as $r \rightarrow \infty$ using the form
\begin{equation}\label{Intro.ansatz}
	g_r \sim \frac{G\,\Gamma(r+\gamma)}{\chi^{r+\gamma}} \qquad \mathrm{as} \quad r \rightarrow \inf\,,
\end{equation}
where $G$, $\gamma$, and $\chi$ are functions that do not depend on $r$ but are free to vary with independent variables (and hence with $z$). The `singulant' $\chi$ equals $0$ at values of $z$, denoted by $z = z_s$, at which the leading-order behavior $g_0$ is singular. This ensures that the late-order ansatz for $g_r$ also has a singularity at $z = z_s$ and that the singularity increases in strength as $r$ increases. \textcolor{black}{In practice, $\gamma$ typically takes constant value, and we will assume this to be the case in the present study.} One can then use the ansatz \eqref{Intro.ansatz} to optimally truncate an asymptotic expansion. The method developed in Olde Daalhuis et al. \cite{Daalhuis1} involves substituting the resulting truncated series expression into the original problem to obtain an equation for the exponentially small remainder term. 

The exponentially small contribution to the asymptotic solution that one obtains using the above method, denoted $g_{\mathrm{exp}}$, generally takes the form in the limit $\delta \rightarrow 0$ given by
\begin{equation}\label{Intro.WKB}
g_{\mathrm{exp}} \sim \mathcal{S}G\e^{-\chi/\delta},
\end{equation} 
where the `Stokes multiplier' $\mathcal{S}$ varies rapidly from $0$ to a nonzero value as one crosses a Stokes curve. This behavior is known as `{Stokes switching}', and it occurs along curves at which the switching exponential is maximally subdominant and hence where the singulant $\chi$ is real and positive \cite{Berry5}. The variation is smooth, and it occurs in a neighborhood of width $\mathcal{O}(\sqrt{\delta})$ that contains the Stokes curve.

%Examining when a Stokes multiplier becomes nonzero provides a simple criterion to determine where the exponentially small contribution to a solution `switches on' and cannot be ignored. One finds the positions of Stokes curves by determining the curves along which the singulant is real and positive. Consequently, exponential asymptotic analysis makes it possible to (1) determine the form of exponentially small contributions to a solution and (2) determine the regions of a solution domain in which the contributions are present. In our study, these contributions take the form of an exponentially small wave train that follows the leading-order soliton, producing nanoptera.

This primary advantages of this approach are that it it reveals the important role played by the Stokes curves in the asymptotic solution, and it does not require the computation of terms beyond the leading-order expression to obtain the form of exponentially small correction terms. The latter of these advantages makes the technique particularly useful for the many nonlinear problems for which obtaining even these low-order correction terms is intractable. See the review article \cite{Boyd1} or monograph \cite{Boyd3} for more details on exponential asymptotics and their application to nonlocal solitary waves, \cite{Berry1,Berry4,Boyd2} for examples of previous studies of exponential asymptotics, and \cite{Chapman1,Daalhuis1} for more details on the particular methodology that we apply in the present paper.

\section{Diatomic FPUT Equation}\label{S.FPUT}

We consider the diatomic FPUT lattice equations given in \eqref{1.FPUT}--\eqref{1.POT}, with $m(j) = m_1$ for $j$ odd, and $m(j) = m_2$ for $j$ even. We express the mass ratio as $\delta = \sqrt{m_2/m_1}$, and set $0 < \delta \ll 1$.

Long waves with small amplitudes in diatomic FPUT lattices were previously studied in \cite{Gaison1} in terms of the offset $r(j,t)$ and the particle velocity $p(j,t)$, defined by
\begin{equation}\label{1.RP}
r(j,t) = x(j+1,t) - x(j,t),\qquad p(j,t) = \dot{x}(j,t),
\end{equation}
\textcolor{black}{giving the governing equations as
\begin{equation}
\dot{r}(j,t) = p(j+1,t) - p(j,t),\qquad m_j\dot{p}(j,t) = F(r(j,t))-F(r(j-1,t)),
\end{equation}
where $m_j = m_1$ if $j$ is odd, and $m_j = m_2$ if $j$ is even.} In \cite{Gaison1}, a long-wave solution was found in terms of a small parameter $\epsilon$ that specifies the amplitude and width of the travelling wave. The long-wave solution was given by
\begin{equation}\label{1.RPsol}
(r(j,t),p(j,t)) = 3\epsilon^2 \sech^2(\beta\epsilon ( j - c_{\epsilon} t) ) (1,-c_0) + \mathcal{O}(\epsilon^{5/2}),
\end{equation}
in the limit that $\epsilon \rightarrow 0$, where
\begin{equation}\label{1.BC}
\beta = \sqrt{\frac{3(m_1^2 + 2 m_1 m_2 + m_2^2)}{2(m_1^2-m_1 m_2 + m_2^2)}}, \qquad c_{\epsilon} = (1 + \epsilon^2)\sqrt{\frac{2}{m_1 + m_2}}.
\end{equation}
We see that small $\epsilon$ corresponds to a long, small-amplitude travelling wave moving at slightly above the speed of sound in the system, denoted $c_0$. By integrating the particle velocities, we obtain
\begin{equation}\label{1.X}
x(j,t) = \frac{3\epsilon}{\beta} \tanh(\beta \epsilon  (j - c_{\epsilon} t)) + \mathcal{O}(\epsilon^{3/2}).
\end{equation}
This approximate travelling wave solution will play an important role in subsequent analysis.

\textcolor{black}{While representing the system in terms of the offset $r$ and particle velocity $p$  is algebraically simpler than the system \eqref{1.FPUT}}, it is convenient for the present analysis to seperate the motion of the heavy and light particles. For clarity in subsequent analysis, we will denote the position of the heavier particles as $y(j,t)$ where $j$ takes odd values, and the lighter particles as $z(j,t)$ where $j$ takes even values. It is always possible to remove the larger mass through non-dimensionalizing the system, so without any loss of generality, we set $m_1 = 1$ and $m_2 = \delta^2$ in all subsequent analysis. \textcolor{black}{We are interested in travelling wave solutions, and hence convert to a moving frame with velocity $c_{\epsilon}$ parameterized by $\xi$, where $\xi = j - c_{\epsilon}t$.} We write $y(j,t) = y(\xi)$ and $z(j,t) = z(\xi)$, and the governing equations become
\begin{align}\label{1.gov1}
c_{\epsilon}^2 y''(\xi) &= F(z(\xi+1) - y(\xi)) - F(y(\xi) - z(\xi-1)),\quad\quad j \,\, \mathrm{odd},\\
\delta^2 c_{\epsilon}^2  z''(\xi) &= F(y(\xi+1) - z(\xi)) - F(z(\xi) - y(\xi-1)),\quad\quad j \,\, \mathrm{even},\label{1.gov2}
\end{align}
where a dash denotes a derivative with respect to $\xi$. 

Before considering the behaviour of this system for small $\delta$, we must determine whether the $\delta = 0$ system contains exponentially small oscillations in the far field due to the discrete nature of the system itself, such as those seen for the discretized Korteweg-de Vries (KdV) equation in \cite{Lustri6}. It is straightforward to see that when $\delta = 0$, the governing equations become
\begin{align}\label{A1.hom1}
c_{\epsilon}^2 y''(\xi) &= F(\tfrac{1}{2}(y(\xi+2) - y(\xi))) - F(\tfrac{1}{2}(y(\xi) - y(\xi-2))), &&\textrm{$j$ odd,}\\
z(\xi) &= \tfrac{1}{2}y(\xi+1) + \tfrac{1}{2}y(\xi-1), && \textrm{$j$ even.}
\end{align}
This shows that the position of the heavy particles is governed by a scaled monoatomic FPUT equation, while the lighter particles occupy the average position of their nearest neighbours. From previous work including \cite{Friesecke1,Friesecke2,Friesecke3}, we know that travelling waves in the monoatomic FPUT lattice are localized solitary waves. Consequently, the long-wave approximation to the $\delta=0$ system does not introduce exponentially small far-field oscillations. This confirms that any far-field oscillations in solutions to the full system \eqref{1.gov1}--\eqref{1.gov2} are introduced by the singular perturbation caused by the small mass ratio parameter $\delta$.

%\textcolor{red}{Before considering the behaviour of this system for small $\delta$, we must determine whether the $\delta = 0$ system contains exponentially small oscillations in the far field due to the discrete nature of the system itself, such as those seen for the discretized Korteweg-de Vries (KdV) equation in \cite{Lustri6}. In Appendix [REF], we apply a similar analysis to \cite{Lustri6} in order to show that corresponding far-field oscillations are not present in this system. This confirms that any far-field oscillations in solutions to the full system \eqref{1.gov1}--\eqref{1.gov2} are introduced by the singular perturbation caused by the small mass ratio parameter $\delta$.}

%\textcolor{black}{Do I have to say something about the relative size of $\epsilon$ and $\delta$? I hope not, but I'd rather not rule it out. The exponentially small waves are induced by the singular perturbation in $\delta$, not the long-wave approximation that is in play with $\epsilon$.}

\section{Series Expansion}\label{S.Series}

We expand the dependent variables as a Taylor series in $\delta$, giving
\begin{equation}\label{2.Ser}
y(\xi) = \sum_{r=0}^{\infty} \delta^{2r} y_r(\xi),\qquad z(\xi) = \sum_{r=0}^{\infty} \delta^{2r} z_r(\xi),
\end{equation}
where the coefficients $y_r$ and $z_r$ depend on the long-wave parameter $\epsilon$, but not the mass ratio $\delta$. We may determine equations for the coefficients by applying these series expressions to \eqref{1.gov1}--\eqref{1.gov2}, and matching in the limit that $\delta \rightarrow 0$. 

\subsection{Leading-order series terms}\label{S.LOT}

At leading order in the limit that $\delta \rightarrow 0$, this gives
\begin{align}\label{2.Hom}
c_{\epsilon}^2 y''_0(\xi) &= F(z_0(\xi+1) - y_0(\xi)) - F(y_0(\xi) - z_0(\xi-1)),\quad\quad j \,\, \mathrm{odd},\\
0 &= F(y_0(\xi+1) - z_0(\xi)) - F(z_0(\xi) - y_0(\xi-1)),\quad\quad j \,\, \mathrm{even}.
\end{align}
This is equivalent to the full diatomic FPUT system \eqref{1.FPUT} with $\delta = 0$. Consequently, \textcolor{black}{we may construct an approximate travelling wave solution by setting} $m_1 = 1$ and $m_2 = 0$ in \eqref{1.RPsol}. This gives the long-wave solution in the limit that $\epsilon \rightarrow 0$ as
\begin{align}\label{2.y0z0}
y_0(\xi) = {\sqrt{6}\,\epsilon} \tanh\left(\sqrt{\tfrac{3}{2}} \epsilon \,\xi \right) + \mathcal{O}(\epsilon^{3/2}),\qquad z_0(\xi) = \tfrac{1}{2}(y(\xi+1) + y(\xi-1)),
\end{align}
\textcolor{black}{where $y_0(\xi)$ describes the behaviour of heavy particles with $j$ odd, and $z_0(\xi)$ describes the behaviour of light particles with $j$ even}. There is a significant difference at this stage between the present analysis of the diatomic FPUT lattice, and the analysis of the diatomic Toda lattice from \cite{Lustri6}; in the analysis of the Toda lattice, the leading order solution is known exactly, whereas for the FPUT lattice, it is approximated up to $\mathcal{O}(\epsilon^{3/2})$. Consequently, we have two small parameters in the system, which introduces an extra source of error into the approximation.

Finally, for the purposes of subsequent analysis, it is important to observe that the expression for $z_0(\xi)$ is singular at $\xi = \xi_{s,N,\pm} $ where
\begin{equation}\label{2.xis}
\xi_{s,N,\pm} = \frac{(2N-1)\pi\i}{\sqrt{6} \epsilon} \pm 1,\qquad N \in \mathbb{Z},
\end{equation}
and that the singular behaviour is given by
\begin{equation}\label{2.z0sing}
z_0(\xi) \sim (\xi - \xi_{s,N,\pm})^{-1} \quad \mathrm{as} \quad \xi \rightarrow \xi_{s,N,\pm},
\end{equation}
for any choice of $N \in \mathbb{Z}$ and sign.

We need only consider the contributions associated with singularities nearest to the real axis, or $N = 1$ and $N = 0$. In all subsequent analysis, we will consider the exponentially small oscillations cased by the singularity in $z_0(\xi)$ at $\xi = \xi_{s,1,-}$. For simplicity of notation, we will denote this particular choice as $\xi = \xi_s$ in subsequent analysis, giving $\xi_s = {\pi\i}/{\sqrt{6} \epsilon}-  1$. 

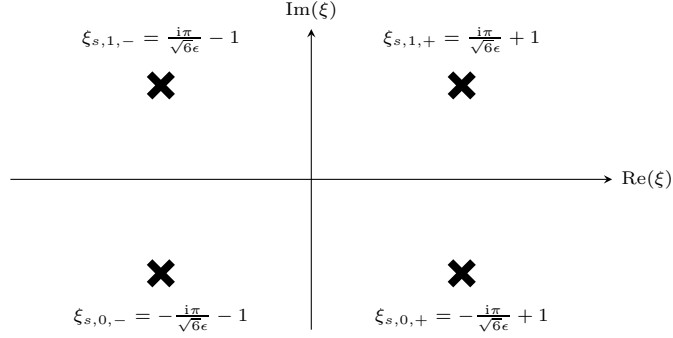
\begin{figure}
\centering
\begin{tikzpicture}
[xscale=1,>=stealth,yscale=1]

%\draw[line width=2pt,black] (-2,1.8) -- (-2,0) node[below] {\scriptsize{$-1$}} -- (3,0);
\draw[line width=3pt] (-2+0.15,1.25+0.15) -- (-2-0.15,1.25-0.15);
\draw[line width=3pt] (-2-0.15,1.25+0.15) -- (-2+0.15,1.25-0.15);
\draw[line width=3pt] (2+0.15,1.25+0.15) -- (2-0.15,1.25-0.15);
\draw[line width=3pt] (2-0.15,1.25+0.15) -- (2+0.15,1.25-0.15);
\draw[line width=3pt] (-2+0.15,-1.25+0.15) -- (-2-0.15,-1.25-0.15);
\draw[line width=3pt] (-2-0.15,-1.25+0.15) -- (-2+0.15,-1.25-0.15);
\draw[line width=3pt] (2+0.15,-1.25+0.15) -- (2-0.15,-1.25-0.15);
\draw[line width=3pt] (2-0.15,-1.25+0.15) -- (2+0.15,-1.25-0.15);

\draw[->] (-4,0) -- (4,0) node[right] {\scriptsize{$\mathrm{Re}(\xi)$}};
\draw[->] (0,-2) -- (0,2) node[above] {\scriptsize{$\mathrm{Im}(\xi)$}};

\node at (-2,1.55) [above] {\scriptsize{$\xi_{s,1,-} =\frac{ \i\pi}{\sqrt{6}\epsilon} - 1$}};
\node at (2,1.55) [above] {\scriptsize{$\xi_{s,1,+} = \frac{ \i\pi}{\sqrt{6}\epsilon} + 1$}};
\node at (-2,-1.55) [below] {\scriptsize{$\xi_{s,0,-} = -\frac{ \i\pi}{\sqrt{6}\epsilon} - 1$}};
\node at (2,-1.55) [below] {\scriptsize{$\xi_{s,0,+} = -\frac{ \i\pi}{\sqrt{6}\epsilon} + 1$}};

\end{tikzpicture}

\caption{Singularities of $z_0(\xi)$ given by \eqref{2.z0sing} that will contribute to the asymptotic form of the far-field oscillations. The subsequent analysis for the oscillations due to the singularity at $\xi = \xi_{s,1,-}$, denoted $\xi_s$, will be shown in detail. The remaining three singularity contributions will be subsequently stated. } 
   \label{F:sing}
\end{figure}

The contributions associated with singularities at $\xi = {\xi}_{s,0,-}$, $\xi_{s,1,+}$, and ${\xi}_{s,0,+}$ will also contribute to the asymptotic form of the far-field oscillations, and the corresponding results will be stated after the conclusion of the detailed analysis for $\xi = \xi_s$. The four singularities are illustrated in Figure \ref{F:sing}.

\subsection{Late-order series terms}\label{S.LateOrder}

The recursion relation is obtained by applying the series expression \eqref{2.Ser} to the governing equations \eqref{1.gov1}--\eqref{1.gov2} and matching orders of $\delta$, giving %\begin{align}
\begin{align}\label{3.recur1}
\nonumber c_{\epsilon}^2 y''_{r}(\xi) = &(z_r(\xi+1) - y_r(\xi))F'(z_0(\xi+1) - y_0(\xi))-(y_r(\xi) - z_r(\xi-1))F'(y_0(\xi) - z_0(\xi-1))\\
\nonumber  &+(z_{1}(\xi+1) - y_{1}(\xi))(z_{r-1}(\xi+1) - y_{r-1}(\xi))F''(z_0(\xi+1)- y_0(\xi))\\
 &-(y_{1}(\xi) - z_{r-1}(\xi-1))(y_{1}(\xi) - z_{r-1}(\xi-1))F''(y_0(\xi) - z_0(\xi-1)) + \ldots,\\
\nonumber c_{\epsilon}^2 z''_{r-1}(\xi) = &(y_r(\xi+1) - z_r(\xi))F'(y_0(\xi+1) - z_0(\xi))-(z_r(\xi) - y_r(\xi-1))F'(z_0(\xi) - y_0(\xi-1))\\
\nonumber  &+(y_{1}(\xi+1) - z_{1}(\xi))(y_{r-1}(\xi+1) - z_{r-1}(\xi))F''(y_0(\xi+1)- z_0(\xi))\\
 &-(z_{1}(\xi) - y_{1}(\xi-1))(z_{r-1}(\xi) - y_{r-1}(\xi-1))F''(z_0(\xi) - y_0(\xi-1)) + \ldots,\label{3.recur2}
\end{align}
\textcolor{black}{where the omitted terms are products containing $y_{r-k}$ and $z_{r-k}$ with $k > 1$. These terms will be smaller than those retained in the limit that $r \rightarrow \infty$, which contain $y_r$, $z_r$, $y_{r-1}$ and $z_{r-1}$. This omission is discussed in more detail at the end of this section, after the late-order ansatz defined in \eqref{3.ansatz} has been applied.}

\textcolor{black}{In principle this recursion relation could be repeatedly applied in order to obtain terms in the series up to arbitrarily large values of $r$, given all previous terms in the series. This process would involve solving the algebraic equation \eqref{3.recur2} for $z_r$, and then solving the differential-difference equation \eqref{3.recur1} to obtain $y_r$,. Repeating this process allows for the calculation of series terms up to arbitrary order; however, this is technically challenging, and it will not reveal the presence of exponentially small oscillations in the far-field, as such oscillations are typically exponentially small in the singularly perturbed limit ($\delta \rightarrow 0$). }

Instead, we must follow \cite{Lustri5} and determine the asymptotic form of the series terms in the limit that $r\rightarrow \infty$, known as the late-order terms. We see that obtaining the value of $y_r$ and $z_r$ for large $r$ requires two differentiations of the term $z_{r-1}$. Hence, we can follow the process devised in Chapman et al. \cite{Chapman1} and apply a factorial-over-power late-order ansatz to approximate these terms as $r \rightarrow \infty$.

The required ansatz takes the form
\begin{equation}\label{3.ansatz}
y_r(\xi) \sim \frac{Y(\xi)\Gamma(2r + \alpha)}{\chi(\xi)^{2r + \alpha}},\qquad z_r(\xi) \sim \frac{Z(\xi)\Gamma(2r + \beta)}{\chi(\xi)^{2r + \beta}},\qquad \mathrm{as} \qquad r \rightarrow \infty,
\end{equation}
\textcolor{black}{where we assume that $\alpha$ and $\beta$ take constant value.\footnote{This assumption may be omitted, as it was in \cite{Chapman1}. In this case, the late-order equation derived by substituting the ansatz into the governing equation, given in this study in \eqref{3.LO}, contains extra terms that are smaller than the leading-order behaviour in \eqref{3.LO} as $r \rightarrow \infty$, but larger than the first correction terms. Matching at this order gives the result that $\gamma$ is constant.}} An identical balancing argument to the equivalent analysis for the diatomic Toda lattice from \cite{Lustri5} gives $\alpha$ = $\beta-2$. 

\textcolor{black}{At this stage we note that for sufficiently large $r$, the terms of the asymptotic series \eqref{2.Ser} diverge in a ``factorial-over-power'' fashion. This divergence is captured by the form of the late-order ansatzes in \eqref{3.ansatz}, where the gamma function grows faster as $r \rightarrow \infty$ than the algebraic decay caused by the increasing power of the singulant in the denominator. Consequently, $y_r \gg y_{r-k}$ and $z_r \gg z_{r-k}$ for $k > 0$ as $r \rightarrow \infty$, with the size being controlled by the argument of the gamma function for asymptotic matching purposes. }

Substituting the late-order terms \eqref{3.ansatz} into the recursion relation \eqref{3.recur2} and keeping only the first two orders in the large $r$ limit gives
\begin{align}
\nonumber  &\frac{c_{\epsilon}^2 (\chi'(\xi))^2 Z(\xi) \Gamma(2r + \alpha)}{\chi(\xi)^{2r+\alpha}} - \frac{2 c_{\epsilon}^2 \chi'(\xi) Z'(\xi) \Gamma(2r + \alpha-1)}{\chi(\xi)^{2r+\alpha-1}} - \frac{ c_{\epsilon}^2 \chi''(\xi) Z(\xi) \Gamma(2r + \alpha-1)}{\chi(\xi)^{2r+\alpha-1}} \\
\nonumber=& -\frac{2 Z(\xi) \Gamma(2r + \alpha)}{\chi(\xi)^{2r + \alpha}}F'(\tfrac{1}{2}(y_0(\xi+1) - y_0(\xi-1)))-\frac{2 Z(\xi) \Gamma(2r + \alpha-2)}{\chi(\xi)^{2r + \alpha-2}}F''(\tfrac{1}{2}(y_0(\xi+1) - y_0(\xi-1))) \\
\nonumber&+\frac{Y(\xi+1) \Gamma(2r + \alpha-2)}{\chi(\xi+1)^{2r + \alpha-2}}(y_{1}(\xi+1) - z_{1}(\xi))F'(\tfrac{1}{2}(y_0(\xi+1) - y_0(\xi-1))) \\
&+\frac{Y(\xi-1)\Gamma(2r + \alpha-2)}{\chi(\xi-1)^{2r + \alpha-2}}(z_{1}(\xi) - y_{1}(\xi-1))F'(\tfrac{1}{2}(y_0(\xi+1) - y_0(\xi-1))) + \ldots,\label{3.LO}
\end{align}
where we use the leading-order relationship between $y_0(\xi)$ and $z_0(\xi)$, and the omitted terms are smaller than those retained as $r \rightarrow \infty$. \textcolor{black}{The leading order terms in this expression grow as $\Gamma(2r + \alpha)/{\chi^{2r+\alpha}}$ in the limit that $r \rightarrow \infty$, and are therefore the same size in this limit as $z_{r+1}$ (noting that $\beta = \alpha - 2$). These factorial-over-power terms of this size are therefore $\mathcal{O}(z_{r+1})$ in the limit that $r \rightarrow \infty$. }

\textcolor{black}{The first correction terms in this expression grow as $\Gamma(2r + \alpha-1)/{\chi^{2r+\alpha-1}}$ in the limit that $r \rightarrow \infty$, and are therefore smaller in this limit as than $z_{r+1}$, but larger than $z_r$. In fact, these terms correspond to the size the late-order ansatz would take for $z_{r+1/2}$, by straightforward substitution. These factorial-over-power terms of this size are therefore $\mathcal{O}(z_{r+1/2})$ in the limit that $r \rightarrow \infty$, even though this is a purely algebraic construction, rather than a term in the asymptotic series \eqref{2.Ser}. }

\textcolor{black}{The terms that were omitted from \eqref{3.LO} all grow as  $\Gamma(2r + \alpha-4)/{\chi^{2r+\alpha-4}}$ in the limit that $r \rightarrow \infty$. Using the notation from before, they are are $\mathcal{O}(z_{r-1})$, and therefore smaller than the terms that were retained in the limit that $r \rightarrow \infty$.}

\subsubsection{Calculating $\chi$}\label{S.Singulant}

\begin{figure}
\centering
\begin{tikzpicture}
[xscale=1,>=stealth,yscale=1]

\draw[line width=2pt,black] (-2,1.8) -- (-2,0) node[below] {\scriptsize{$-1$}} -- (3,0);
\draw[line width=3pt] (-2+0.15,1.8+0.15) -- (-2-0.15,1.8-0.15);
\draw[line width=3pt] (-2-0.15,1.8+0.15) -- (-2+0.15,1.8-0.15);
\fill[black] (3,0) circle (0.1);

\draw[->] (-4,0) -- (4,0) node[right] {\scriptsize{$\mathrm{Re}(s)$}};
\draw[->] (0,-1) -- (0,2.5) node[above] {\scriptsize{$\mathrm{Im}(s)$}};

\node at (-2,2) [above] {\scriptsize{$\xi_s = \frac{ \i\pi}{\sqrt{6}\epsilon} - 1$}};
\node at (3,-0.05) [below] {\scriptsize{$\xi$}};
\node at (-2,0.9) [left] {\scriptsize{$\mathcal{C}_1$}};
\node at (1,0) [below] {\scriptsize{$\mathcal{C}_2$}};

\end{tikzpicture}

\caption{The integral contour for \eqref{3.chiint}, which connects the singularity location $s = \xi_s$ with $s = \xi$. The location of the singularity is denoted by a cross, while the contour is a thick black line. The integral contour may be divided into a vertical component $\mathcal{C}_1$ and a horizontal component $\mathcal{C}_2$. The integral contribution along $\mathcal{C}_1$ is real, and the integral contribution along $\mathcal{C}_2$ is imaginary; this implies that $\mathrm{Re}(\chi)$ is constant for real-valued $\xi$.} 
   \label{F:Contours}
\end{figure}
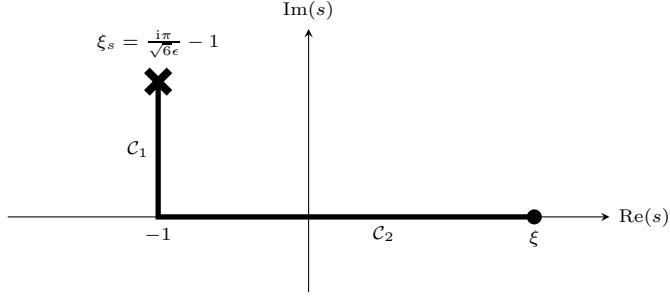

Matching \eqref{3.LO} at $\mathcal{O}(z_{r+1})$ as $r \rightarrow \infty$, we find that
\begin{equation}\label{3.chieq}
c_{\epsilon}^2 (\chi')^2 = -2 (1 +  y_0(\xi+1) - y_0(\xi-1)).
\end{equation}
Integrating this, and recalling that $\chi = 0$ \textcolor{black}{at} the singularity location $\xi = \xi_s$ gives
\begin{equation}\label{3.chiint}
\chi = \pm \frac{\i\sqrt{2}}{c_{\epsilon}} \int_{\xi_{s}}^{\xi} \sqrt{1 + y_0(s+1) - y_0(s-1)} \d s.
\end{equation}
The integral contour is depicted in Figure \ref{F:Contours}. While any contour may be chosen, it is helpful to divide the contour into a vertical component $\mathcal{C}_1$ and a horizontal component $\mathcal{C}_2$. The real contribution to this integral arises by integrating down $\mathcal{C}_1$, while the imaginary contribution is caused by integrating along $\mathcal{C}_2$. 

Consequently, this allows us to obtain the real and imaginary parts of the singulant,
\begin{align}\label{3.chiintrealimag}
\mathrm{Re}(\chi) = \pm\frac{\i\sqrt{2}}{c_{\epsilon}} \int_{\mathcal{C}_1} \sqrt{1 + y_0(s+1) - y_0(s-1)} \d s,\qquad 
\mathrm{Im}(\chi) = \pm\frac{\sqrt{2}}{c_{\epsilon}}\int_{\mathcal{C}_2} \sqrt{1 + y_0(s+1) - y_0(s-1)} \d s.
\end{align}
We note that $\mathrm{Re}(\chi)$ is constant for real $\xi$, which causes the far field oscillations to have constant amplitude. 

%We will eventually determine that the far-field oscillations have the exponential form derived later in \eqref{4.wkb}, and the amplitude of the far-field oscillations is controlled by $\mathrm{Re}(\chi)$. Consequently, the far-field oscillations have constant amplitude.

%In order to determine the sign choice, we find an asymptotic expression in the $\epsilon \rightarrow 0$ limit for $\mathrm{Re}(\chi)$, finding
%\begin{equation}\label{3.chiintreal}
%\mathrm{Re}(\chi) = \pm\frac{\i\sqrt{2}}{c_{\epsilon}} \int_{\mathcal{C}_1} \sqrt{1 + y_0(\xi+1) - y_0(\xi-1)} \d s \sim \pm\frac{\pi}{\sqrt{6}\epsilon} \quad \mathrm{as} \quad \epsilon \rightarrow 0.
%\end{equation}
For Stokes switching to occur, we recall from Section \ref{S.AsympIntro} that $\mathrm{Re}(\chi)>0$, which corresponds to the positive choice of sign. Hence, we restrict our subsequent analysis to this sign choice.

%Consequently, numerical evaluation of the full integral expression \eqref{3.chiint} will be used for subsequent calculations, with the positive sign chosen.

%\subsubsection{Asymptotic behaviour of $\chi$}
%
%Far from the travelling wave front, we know that $y_0(\xi+1)- y_0(\xi-1) \rightarrow 0$. Consequently, as $|\xi| \rightarrow \infty$, we have
%\begin{equation}
%\chi \sim \frac{ \i\sqrt{2}}{c_{\epsilon}} (\xi - \xi_{s}),
%\end{equation}

It will be important in the subsequent analysis to know the local behaviour of the singulant near the singular point. It is straightforward to show by direct computation that
\begin{equation}
c_{\epsilon}^2 (\chi')^2  \sim  -{4}(\xi - \xi_{s})^{-1} \quad \mathrm{as} \quad \xi \rightarrow \xi_{s}.\label{3.chiloc1}
\end{equation}
\textcolor{black}{The explicit local expansions for $y_0(\xi + 1)$ and $y_0(\xi-1)$ are given in \eqref{A.y0p1} and \eqref{A.y0m1}. The dominant contribution to the right-hand side of \eqref{3.chiloc1} comes from the leading-order local behaviour of $y_0(\xi+1)$, which is independent of $\epsilon$.} This local expression may now be rearranged to give
\begin{equation}\label{3.chiloc}
\chi \sim \frac{4\i}{c_{\epsilon}} (\xi - \xi_{s})^{1/2} \quad \mathrm{as} \quad \xi \rightarrow \xi_{s}.
\end{equation}

%We note that the calculations in this section are not directly dependent on the long-wave parameter $\epsilon$ as they use the exact leading-order solution $y_0(\xi)$, although the position of the singularity $\xi_s$ does depend on $\epsilon$. In order to compute the integral, the approximation for $y_0(\xi)$ given in \eqref{2.y0z0} may be used. This introduces an $\epsilon$-dependent asymptotic error into the integral expression \eqref{3.chiintrealimag}, which must decay as $\epsilon \rightarrow 0$.

\subsubsection{Calculating $Z$ and $\alpha$}\label{S.ZAlpha}

Matching \eqref{3.LO} at $\mathcal{O}(z_{r-1/2})$, after some simplification,
\begin{equation}\label{3.Zeq}
2 \chi' Z' + \chi'' Z  = 0.
\end{equation}
This expression may be integrated to give
\begin{equation}\label{3.Z}
Z = \frac{\Lambda}{\sqrt{\chi'(\xi)}},
\end{equation}
where $\Lambda$ is an arbitrary constant that one may determine by considering an inner expansion of the solution in the neighbourhood of $\xi = \xi_s$, and matching the outer limit of this expansion with the inner limit of the late-order ansatz \eqref{3.ansatz}. Performing this inner analysis requires the value of $\alpha$ from \eqref{3.ansatz}.

%We also have that 
%\begin{equation}
%V = \frac{\Lambda}{\sqrt{\chi'(\xi)}},
%\end{equation}
%which gives
To determine $\alpha$, we use \eqref{3.chiloc} and \eqref{3.Z} to find a local expression for $Z$. Direct computation gives
\begin{equation}\label{3.Zloc}
Z \sim \frac{\Lambda c_{\eps}^{1/2}}{2^{1/2}}(\xi-\xi_s)^{1/4} \quad \mathrm{as} \quad \xi \rightarrow \xi_s.
\end{equation}
By comparing this expression with the local form of the late-order ansatz, we see that 
\begin{equation}\label{3.zrloc}
z_r(\xi) \sim \frac{\Lambda c_{\eps}^{1/2}(\xi-\xi_s)^{1/4}\Gamma(2r + \gamma)}{2^{1/2}(4\i(\xi-\xi_s)^{1/2}/c_{\epsilon})^{2r+\alpha}}\quad \mathrm{as} \quad r\rightarrow\infty,\, \xi \rightarrow \xi_s.
\end{equation}
In order for this equation to be consistent with the initial behaviour, which has singularity strength of one at $\xi = \xi_s$, we require that $\alpha/2 - 1/4 = 1$, which gives $\alpha = 5/2$.

Using this value of $\alpha$, as well as local expressions for $\chi$ and $Z$ from \eqref{3.chiloc} and \eqref{3.Zloc} respectively, we may perform a local analysis to determine $z_r$ in the neighbourhood of $\xi = \xi_s$, and apply asymptotic matching in order to determine the unknown $\Lambda$. The detailed local analysis is presented in Appendix \ref{A.Lambda}, and yields $\Lambda = {8\sqrt{\pi\i}}/{c_{\epsilon}}$.

We have now fully determined the late-order ansatz form for $z_r$, given in \eqref{3.ansatz}, and know that
\begin{equation}
z_r(\xi) \sim \frac{8\sqrt{\pi\i}\Gamma(2r + 5/2)}{c_{\epsilon}\sqrt{\chi'(\xi)}\chi(\xi)^{2r + 5/2}} \quad \mathrm{as} \quad r \rightarrow \infty,
\end{equation}
where $\chi$ is given by \eqref{3.chiint} with the positive choice of sign. 

%It is straightforward to work out corresponding expressions associated with each of the three remaining singularities in $z_0(\xi)$. 

\section{Exponential Asymptotics}\label{S.ExpAsymp}

The main idea of exponential asymptotics is that one can truncate a divergent asymptotic series optimally at some term number $N_{\mathrm{opt}}$ such that the remainder is exponentially small in size. The truncated series can be expressed as
\begin{equation}\label{4.series}
	y(\xi)= \sum_{r = 0}^{N_{\mathrm{opt}}-1}\delta^{2r} y_r(\xi) + y_{\mathrm{exp}}(\xi)\,, \qquad z(\xi)= \sum_{j=0}^{N_{\mathrm{opt}}-1} \delta^{2r} z_r(\xi) + z_{\mathrm{exp}}(\xi)\,.
\end{equation}
where $y_{\mathrm{exp}}$ and $z_{\mathrm{exp}}$ are exponentially small in the limit $\delta \rightarrow 0$.

\subsection{Stokes Structure}\label{S.StokesStruct}

We recall that there are four exponentially small asymptotic contributions, associated with the singularities depicted in Figure \ref{F:sing}. We denote the corresponding singulants as $\chi_{1,-}$, $\chi_{0,-}$, $\chi_{1,+}$ and $\chi_{0,+}$.

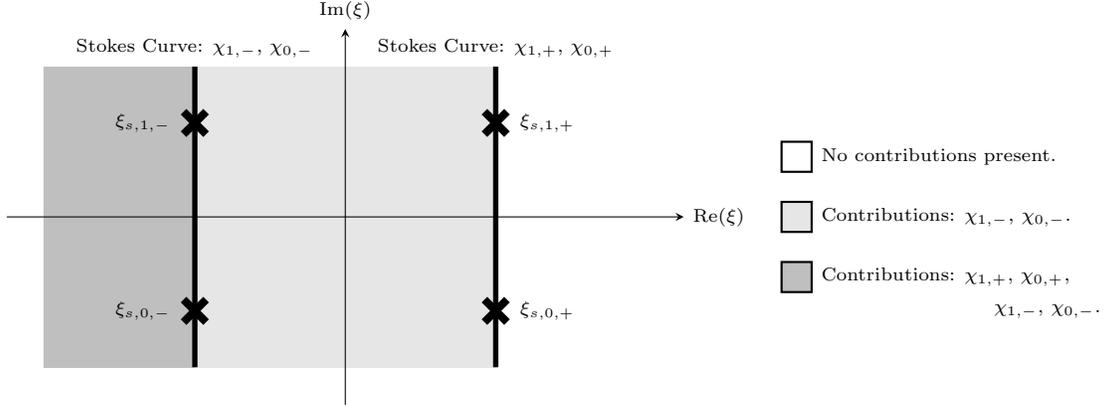
\begin{figure}
\centering
\begin{tikzpicture}
[xscale=1,>=stealth,yscale=1]

\fill[black!10] (2,-2) -- (-2,-2) -- (-2,2) -- (2,2) -- cycle;
\fill[black!25] (-2,-2) -- (-4,-2) -- (-4,2) -- (-2,2) -- cycle;

\draw[line width=2pt] (2,-2) -- (2,2);
\draw[line width=2pt] (-2,-2) -- (-2,2);

%\draw[line width=2pt,black] (-2,1.8) -- (-2,0) node[below] {\scriptsize{$-1$}} -- (3,0);
\draw[line width=3pt] (-2+0.15,1.25+0.15) -- (-2-0.15,1.25-0.15);
\draw[line width=3pt] (-2-0.15,1.25+0.15) -- (-2+0.15,1.25-0.15);
\draw[line width=3pt] (2+0.15,1.25+0.15) -- (2-0.15,1.25-0.15);
\draw[line width=3pt] (2-0.15,1.25+0.15) -- (2+0.15,1.25-0.15);
\draw[line width=3pt] (-2+0.15,-1.25+0.15) -- (-2-0.15,-1.25-0.15);
\draw[line width=3pt] (-2-0.15,-1.25+0.15) -- (-2+0.15,-1.25-0.15);
\draw[line width=3pt] (2+0.15,-1.25+0.15) -- (2-0.15,-1.25-0.15);
\draw[line width=3pt] (2-0.15,-1.25+0.15) -- (2+0.15,-1.25-0.15);

\draw[->] (-4.5,0) -- (4.5,0) node[right] {\scriptsize{$\mathrm{Re}(\xi)$}};
\draw[->] (0,-2.5) -- (0,2.5) node[above] {\scriptsize{$\mathrm{Im}(\xi)$}};

\node at (-2.2,1.25) [ left] {\scriptsize{$\xi_{s,1,-}$}};
\node at (2.2,1.25) [ right] {\scriptsize{$\xi_{s,1,+}$}};
\node at (-2.2,-1.25) [ left] {\scriptsize{$\xi_{s,0,-}$}};
\node at (2.2,-1.25) [ right] {\scriptsize{$\xi_{s,0,+} $}};

\node at (-2,2) [above] {\scriptsize{Stokes Curve: $\chi_{1,-}$, $\chi_{0,-}$}};
\node at (2,2) [above] {\scriptsize{Stokes Curve: $\chi_{1,+}$, $\chi_{0,+}$}};

\draw[thick] (6.2,1.5-0.7) -- (6.2,1.7-0.7) -- (5.8,1.7-0.7) -- (5.8,1.3-0.7) -- (6.2,1.3-0.7) -- (6.2,1.5-0.7) node [right] {\scriptsize{No contributions present.}};
\fill[black!10] (6.2,0.5+0.2-0.7) -- (6.2,0.7+0.2-0.7) -- (5.8,0.7+0.2-0.7) -- (5.8,0.3+0.2-0.7) -- (6.2,0.3+0.2-0.7) -- (6.2,0.5+0.2-0.7);
\draw[thick] (6.2,0.5+0.2-0.7) -- (6.2,0.7+0.2-0.7) -- (5.8,0.7+0.2-0.7) -- (5.8,0.3+0.2-0.7) -- (6.2,0.3+0.2-0.7) -- (6.2,0.5+0.2-0.7) node [right] {\scriptsize{Contributions: $\chi_{1,-}$, $\chi_{0,-}.$}};
\fill[black!25] (6.2,0.5+0.2-0.8-0.7) -- (6.2,0.7+0.2-0.8-0.7) -- (5.8,0.7+0.2-0.8-0.7) -- (5.8,0.3+0.2-0.8-0.7) -- (6.2,0.3+0.2-0.8-0.7) -- (6.2,0.5+0.2-0.8-0.7);
\draw[thick] (6.2,0.5+0.2-0.8-0.7) -- (6.2,0.7+0.2-0.8-0.7) -- (5.8,0.7+0.2-0.8-0.7) -- (5.8,0.3+0.2-0.8-0.7) -- (6.2,0.3+0.2-0.8-0.7) -- (6.2,0.5+0.2-0.8-0.7) node [right] {\scriptsize{Contributions: $\chi_{1,+}$, $\chi_{0,+},$}};
\node at (8.5,0.5+0.2-0.8-0.45-0.7) [right] [right] {\scriptsize{$\chi_{1,-}$, $\chi_{0,-}$.}};

\end{tikzpicture}

\caption{Stokes structure for $z(\xi)$. The Stokes curves are represented as black lines, which originate at the singularities of $z_0(\xi)$ (denoted by crosses). The wave front is located at $\xi = 0$, while $\xi > 0$ corresponds to the region ahead of the travelling wave. \textcolor{black}{We expect the solution to be undisturbed far ahead of the wave, as $\xi \rightarrow \infty$. Exponentially small oscillations are switched on as each Stokes curve is crossed from right to left.} In the light gray region, oscillatory contributions caused by the singularities at $\xi = \xi_{s,1,-}$ and $\xi = \xi_{s,0,-}$ are present in the asymptotic solution. In the darker gray region far behind the wave front, oscillatory contributions caused by all four singularities are present in the asymptotic solution. }
   \label{F:Stokes}
\end{figure}

Recall from Section \ref{S.AsympIntro} that Stokes curves are curves in the complex plane that correspond to $\mathrm{Im}(\chi) = 0$ and $\mathrm{Re}(\chi) > 0$. This means that the Stokes structure may be fully determined from the form of \eqref{3.chiint}, as well as the corresponding singulant equation for the three remaining singularity contributions of interest. In each case, the Stokes curves extend vertically from the singularities of $z_0(\xi)$. This is depicted in Figure \ref{F:Stokes}.

We prescribe that the behaviour preceding the wave front is undisturbed, and hence conclude that $z_{\mathrm{exp}}$ is zero as $\xi \rightarrow -\infty$. We therefore conclude that the four exponentially small contributions to $z_{\mathrm{exp}}$ are present to the right of the corresponding Stokes curves, and therefore are only present in the wake of the leading-order travelling wave.

\subsection{Remainder Calculations}\label{S.Remainder}

Finally, we may determine the form of the contributions that are switched on as the Stokes curves are crossed. In order to accomplish this, we truncate the asymptotic series after $N$ terms, giving
\begin{equation}\label{4.series2}
	y(\xi)= \sum_{r = 0}^{{N}-1}\delta^{2r} y_r(\xi) + S_N(\xi)\,, \qquad z(\xi)= \sum_{j=0}^{{N}-1} \delta^{2r} z_r(\xi) + R_N(\xi)\,,
\end{equation}
where $S_N$ and $R_N$ are the remainders obtained by truncating the series after $N$ terms. If $N$ is chosen optimally, these remainder terms are exponentially small \cite{Berry1}. We may then analyse these remainder terms in a neighbourhood of the Stokes curves in order to determine the exponentially small asymptotic contributions to the solution behaviour obtained as the Stokes curves are crossed.

In many places, the subsequent analysis is nearly identical to \cite{Lustri5}, and hence we omit some intermediate technical details. The reader may refer to \cite{Lustri5} for a more detailed presentation of the asymptotic analysis contained in this section, or more general descriptions of the technique described in \cite{Chapman1,Daalhuis1}.

To optimally truncate the series, we apply the heuristic described in \cite{Boyd1}, and determine the optimal truncation  point by finding the point at which consecutive terms in the series are equal in size. This heuristic gives $N_{\mathrm{opt}} = {|\chi|}/{2{\delta}} + \omega$, where we choose $\omega \in [0,1)$ in a way that ensures that $N_{\mathrm{opt}}$ is an integer. We note that $N_{\mathrm{opt}} \rightarrow \infty$ as $\delta \rightarrow 0$, as expected.
%\begin{equation}
%	\left|\frac{\delta^{2N+2} z_{N+1}(\xi)}{\delta^{2N} z_{N}(\xi)}\right| \sim 1 \qquad \mathrm{as} \qquad \delta \rightarrow 0\,.
%\end{equation}
%Using the general form of the ansatz, we obtain $N \sim |\chi|/2{\delta}$. Consequently, we set the optimal truncation point to be
%\begin{equation}
%	N = \frac{|\chi|}{2{\delta}} + \omega\,,
%\end{equation}

We now apply the truncated series expression to the governing equation. Using the recursion relation to simplify the result, we obtain as $\delta \rightarrow 0$ that
\begin{align}\label{4.truncgov1}
	c_{\eps}^2 S_N''(\xi) \sim &(R_N(\xi+1) + R_N(\xi-1))(1 + 2y_0(\xi+2) - 2y_0(\xi))\\
	& -(R_N(\xi+1) + R_N(\xi-1))(1 + y_0(\xi) - y_0(\xi-2)) + \ldots,\nonumber\\
	c_{\eps}^2 \delta^2  R_N''(\xi)& + 2(1+2y_0(\xi+1)-2y_0(\xi-1)) R_N(\xi) \sim -  c^2 \delta^{2N} z''_{N-1}(\xi) + \ldots\,,\label{4.truncgov2}
\end{align}
where the terms that we omitted are smaller in magnitude than those that we retained by a factor of $\delta$ or more in the limit that $\delta \rightarrow 0$. Throughout the remainder of this section, the asymptotic limit under consideration is $\delta \rightarrow 0$, which will be omitted for simplicity.

Assuming the truncation point occurs after a sufficiently large number of terms, we may apply the expression for the singulant and the late-order term ansatz to \eqref{4.truncgov2} to obtain
\begin{equation}\label{4.rneq}
	 \delta^2 R''_N(\xi) -  \chi'(\xi)^2 R_N(\xi) \sim -\frac{\delta^{2N} \chi'(\xi)^2 V(\xi) \Gamma(2N +5/2)}{\chi(\xi)^{2N+5/2}}\,.
\end{equation}
%Away from the  Stokes curve, the right-hand side is small. This gives
%\begin{equation}
%	\delta^2 R''_N(\xi) - \chi'(\xi)^2 R_N(\xi) \sim 0,
%\end{equation}
%and 
Outside of a region in the neighbourhood of the Stokes curve, the right-hand side of this expression is exponentially small. A Green-Liouville (or WKBJ) analysis outside of this neighbourhood shows that away from Stokes curves, the remainder takes the form $R_N \sim C V \e^{-\chi/{\delta}}$ as $\delta \rightarrow 0$, where $C$ is an arbitrary constant. In order to capture the variation in the neighbourhood of the Stokes curve, we write
\begin{equation}\label{4.wkb}
	R_N \sim A(\xi) V(\xi)\e^{-\chi(\xi)/{\delta}},% \qquad \mathrm{as} \qquad \delta \rightarrow 0\,,
\end{equation}
where $A(\xi)$ is a Stokes switching parameter that varies rapidly near the Stokes curve. Applying this expression to \eqref{4.rneq} and simplifying gives
%\begin{align}
%	\delta^2\left[\frac{A V (\chi')^2}{\delta^2} - \frac{2A' V \chi'}{{\delta}} - \frac{2 A V'\chi'}{{\delta}} - \frac{A V \chi''}{{\delta}}\right]\e^{-\chi/{\delta}} - (\chi')^2 A V \e^{-\chi/{\delta}} &\sim -\frac{\delta^{2N} (\chi')^2 V \Gamma(2N+5/2)}{\chi^{2N+5/2}}\,.
%\end{align}
%This simplifies to give
\begin{align}
	-2A'\chi'e^{-\chi/{\delta}}  \sim -\frac{\delta^{2N-1} (\chi')^2 \Gamma(2N+5/2)}{\chi^{2N+5/2}},% \qquad \mathrm{as} \qquad \delta \rightarrow 0\,.
\end{align}
We write this using $\chi$ as an independent variable. Noting that $A'(\xi) = \chi'(\xi)\frac{\d A}{\d \chi}$ and rearranging gives
\begin{align}
	\diff{A}{\chi} \sim \frac{\delta^{2N-1} \Gamma(2N+5/2)}{2\chi^{2N+5/2}} \e^{\chi/{\delta}} \,.
\end{align}
We define polar coordinates ($\chi = r\e^{ \i \theta}$) and consider only variation that occurs in the angular direction. Applying the optimal truncation $N = N_{\mathrm{opt}}$, and using an asymptotic expansion for the gamma function \cite{DLMF} subsequently gives
%We thus see that
%\begin{equation}
%	\chi = r\e^{ \i \theta}\,, \qquad \diff{}{\chi} = -\frac{ \i \e^{- \i \theta}}{r}\diff{}{\theta}\,.
%\end{equation}
%The optimal truncation then gives
%\begin{align}
%	\diff{A}{\theta} \sim  \i  r \e^{ \i \theta} \frac{\delta^{r/2{\delta}+2\omega-1} \Gamma(r/{\delta} + 2\omega+5/2)}{2(r \e^{ \i \theta})^{r/\sqrt{\delta}+2\omega+5/2}} \e^{r\e^{ \i \theta}/{\delta}}\,.
%\end{align}
%We now use an asymptotic expansion of the gamma function to obtain
\begin{equation}
	\diff{A}{\theta} \sim  \i \delta \sqrt{\frac{\pi r}{2}}\exp\left(\frac{r}{\sqrt{\delta}}(\e^{ \i \theta}-1) - \frac{ \i \theta r}{\sqrt{\delta}} + \i \theta(1-5/2-2\omega)\right)\,.
\end{equation}
The right-hand side of this expression is exponentially small in $\delta$, except in the neighborhood of $\theta = 0$. We therefore define an inner region $\theta = \delta^{1/2}\bar{\theta}$ and thereby find that
\begin{equation}
	\diff{A}{\bar{\theta}} \sim  \i \delta^{3/2} \sqrt{\frac{\pi r}{2}}\e^{-r\bar{\theta}^2/2}\,.
\end{equation}
Consequently, by integration, we see that the behavior as the Stokes curve is crossed is 
\begin{equation}
	A \sim  \i \delta^{3/2} \sqrt{\frac{\pi r}{2}}  \int_{-\inf}^{\bar{\theta}}\e^{-r s^2/2} \d s \quad \mathrm{as} \quad \delta \rightarrow 0\,.
\end{equation}
Therefore, converting back to outer coordinates, as the Stokes curve is crossed from ${\theta} < 0$ to ${\theta} > 0$, we find that $\left[A\right]_-^+ \sim \pi \i \delta^{3/2}$ as $\delta \rightarrow 0$, where we use the notation $[f]_-^+$ to describe the change in a function $f$ as the Stokes curve is crossed from $\theta<0$ to $\theta>0$. 
%Therefore, the total exponentially small contribution is given in the $\delta \rightarrow 0$ limit by
%\begin{equation}
%	\left[R_N\right]_-^+ \sim \pi \i \delta^{3/2} V \e^{-\chi/\delta} + \mathrm{c.c.}\,,
%\end{equation}
%where `c.c.' stands for the complex conjugate. 
Because $R_N(\xi)$ must be zero ahead of the solitary wave\textcolor{black}{ (i.e., when $\mathrm{Im}(\chi) > 0$), we find from \eqref{4.wkb} that behind the wave front, the exponentially small contribution is given in $\delta \rightarrow 0$ limit by
\begin{equation}
	R_N \sim -\frac{\pi\mathcal{S}(\xi)\Lambda \i \delta^{3/2}}{\sqrt{\chi'(\xi)}}  \e^{-\chi(\xi)/\delta} \quad \mathrm{as}\quad \delta \rightarrow 0.
\end{equation}
where $\mathcal{S}$ varies rapidly from $0$ to $1$ in the neighborhood of the Stokes curve as it is crossed.}

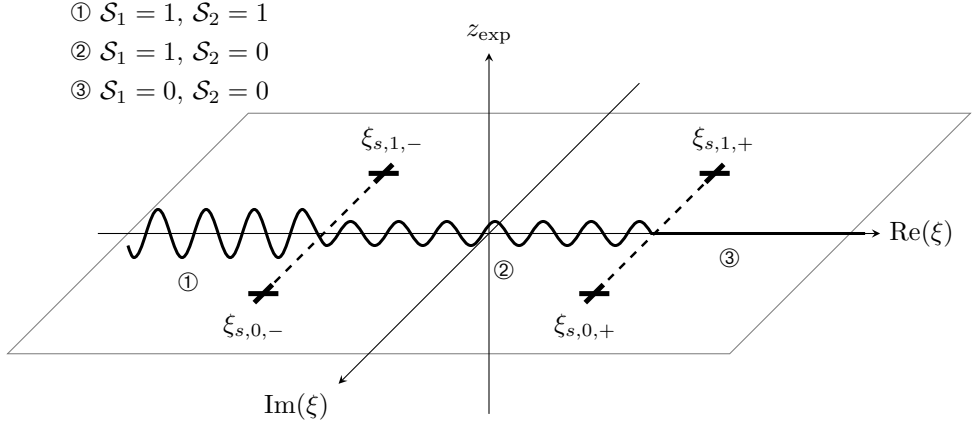
\begin{figure}
\centering
\begin{tikzpicture}
[xscale=0.8,>=stealth,yscale=0.8]

\draw[gray] (-4,2) -- (8,2) -- (4,-2) -- (-8,-2) -- cycle;
\draw[->] (2.5,2.5) -- (-2.5,-2.5) node[below left] {$\mathrm{Im}(\xi)$};
\draw[->] (-6.5,0) -- (6.5,0) node[right] {$\mathrm{Re}(\xi)$};
\draw[->] (0,-3) -- (0,3) node[above] {$z_{\mathrm{exp}}$};

\draw[line width=0.67mm] (3+0.15+0.75,1+0.15) node[above] {$\xi_{s,1,+}$}-- (3-0.15+0.75,1-0.15);
\draw[line width=0.67mm] (3+0.25+0.75,1) -- (3-0.25+0.75,1);

\draw[line width=0.67mm] (1+0.15+0.75,-1+0.15) -- (1-0.15+0.75,-1-0.15) node[below] {$\xi_{s,0,+}$} ;
\draw[line width=0.67mm] (1+0.25+0.75,-1) -- (1-0.25+0.75,-1);
\draw[line width=0.33mm,dashed] (3+0.75,1) -- (1+0.75,-1);

\draw[line width=0.67mm] (3+0.15-4-0.75,1+0.15) node[above] {$\xi_{s,1,-}$}-- (3-0.15-4-0.75,1-0.15);
\draw[line width=0.67mm] (3+0.25-4-0.75,1) -- (3-0.25-4-0.75,1);

\draw[line width=0.67mm] (1+0.15-4-0.75,-1+0.15) -- (1-0.15-4-0.75,-1-0.15)node[below] {$\xi_{s,0,-}$} ;
\draw[line width=0.67mm] (1+0.25-4-0.75,-1) -- (1-0.25-4-0.75,-1);
\draw[line width=0.33mm,dashed] (1-4-0.75,-1) -- (3-4-0.75,1);

\draw node at (-5,-0.8) {{\ding{192}}};
\draw node at (4,-0.4) {{\ding{194}}};
\draw node at (0.25,-0.6) {{\ding{193}}};

%\draw[line width=0.4mm] (-6.25,0) -- (-2.7,0) sin (-2.5,0.2) cos (-2.3,0) sin (-2.1,-0.2) cos (-1.9,0) sin (-1.7,0.2) cos (-1.5,0) sin (-1.3,-0.2) cos (-1.1,0) sin (-0.9,0.2) cos (-0.7,0) sin (-0.5,-0.2) cos (-0.3,0) sin (-0.1,0.2) cos (0.1,0) sin (0.3,-0.2) cos (0.5,0) sin (0.7,0.2) cos (0.9,0) sin (1.1,-0.2) cos (1.3,0) sin (1.5,0.2) cos (1.7,0) sin (1.9,-0.2) cos (2.1,0) sin (2.3,0.2) cos (2.5,0) sin(2.7,-0.2) .. controls (2.8,-0.2) and (3,0.4) .. (3.1,0.4) cos (3.3,0) sin (3.5,-0.4) cos (3.7,0) sin (3.9,0.4) cos (4.1,0) sin (4.3,-0.4) cos (4.5,0) sin (4.7,0.4) cos (4.9,0) sin (5.1,-0.4) cos (5.3,0) sin (5.5,0.4) cos (5.7,0) sin (5.9,-0.4) cos (6,-0.2);
\draw[line width=0.4mm] (6.25,0) -- (2.7,0) sin (2.5,0.2) cos (2.3,0) sin (2.1,-0.2) cos (1.9,0) sin (1.7,0.2) cos (1.5,0) sin (1.3,-0.2) cos (1.1,0) sin (0.9,0.2) cos (0.7,0) sin (0.5,-0.2) cos (0.3,0) sin (0.1,0.2) cos (-0.1,0) sin (-0.3,-0.2) cos (-0.5,0) sin (-0.7,0.2) cos (-0.9,0) sin (-1.1,-0.2) cos (-1.3,0) sin (-1.5,0.2) cos (-1.7,0) sin (-1.9,-0.2) cos (-2.1,0) sin (-2.3,0.2) cos (-2.5,0) sin(-2.7,-0.2) .. controls (-2.8,-0.2) and (-3,0.4) .. (-3.1,0.4) cos (-3.3,0) sin (-3.5,-0.4) cos (-3.7,0) sin (-3.9,0.4) cos (-4.1,0) sin (-4.3,-0.4) cos (-4.5,0) sin (-4.7,0.4) cos (-4.9,0) sin (-5.1,-0.4) cos (-5.3,0) sin (-5.5,0.4) cos (-5.7,0) sin (-5.9,-0.4) cos (-6,-0.2);

\node at (-3.5,3.65) [left] {\ding{192} $\mathcal{S}_1 = 1$, $\mathcal{S}_2=1$};
\node at (-3.5,3) [left] {\ding{193} $\mathcal{S}_1 = 1$, $\mathcal{S}_2=0$};
\node at (-3.5,2.35) [left] {\ding{194} $\mathcal{S}_1 = 0$, $\mathcal{S}_2=0$};

\end{tikzpicture}

\caption{Behavior of the asymptotic coefficients as the Stokes curves are crossed and a schematic illustration of the associated exponentially small contribution. \textcolor{black}{Recall that if $\xi \in \mathbb{R}$, then $\xi > 0$ corresponds the undisturbed region ahead of the wave; in contrast, $\xi < 0$ describes the region in the wake of the wave. In the first region, both of the multipliers are zero, and there are no oscillations. As one crosses the Stokes curve along $\mathrm{Re}(\xi) = 1$, the coefficient $\mathcal{S}_2$ changes rapidly from $0$ to $1$. As one crosses the Stokes curve across $\mathrm{Re}(\xi) = -1$, the coefficient $\mathcal{S}_1$ varies rapidly from $0$ to $1$, ensuring that all four contributions are present in the third region, far behind the wave front.}}\label{F:Stokes2}
\end{figure}

We recall that there are four relevant exponentially small contributions, associated with each of the four singularities in Figure \ref{F:sing}. The contributions associated with $\chi_{0,-}$ and $\chi_{0,+}$ are the complex conjugates of those associated with $\chi_{1,-}$ and $\chi_{1,+}$ respectively. We consequently find that the exponentially small contribution to the asymptotic behaviour in the wake of the leading-order solitary wave is given by
\begin{equation}\label{4.zexp0}
z_{\mathrm{exp}} \sim \left[-\frac{\pi\mathcal{S}_1(\xi)\Lambda \i \delta^{3/2}}{\sqrt{\chi'_{1,-}(\xi)}}  \e^{-\chi_{1,-}(\xi)/\delta} - \frac{\pi\mathcal{S}_2(\xi)\Lambda \i \delta^{3/2}}{\sqrt{\chi'_{1,+}(\xi)}}  \e^{-\chi_{1,+}(\xi)/\delta}\right] + \mathrm{c.c.} \quad \mathrm{as} \quad \delta \rightarrow 0,
\end{equation}
where $\mathcal{S}_1$ varies from zero to one in the neighbourhood of the Stokes curve that follows $\mathrm{Re}(\xi) = -1$ \textcolor{black}{as it is crossed from right to left,} $\mathcal{S}_2$ varies from zero to one in the neighbourhood of the Stokes curve that follows $\mathrm{Re}(\xi) = -1$ \textcolor{black}{as it is crossed from right to left}, and c.c. denotes the complex conjugate. The behaviour of these coefficients is illustrated in Figure \ref{F:Stokes2}.

Noting that $\chi_{1,-}'(\xi) = \chi_{1,+}'(\xi)$, we simplify this to give
\begin{equation}\label{4.zexp}
z_{\mathrm{exp}}(\xi) \sim -\frac{\pi\Lambda \i \delta^{3/2}}{\sqrt{\chi'_{1,-}(\xi)}} \left[ \mathcal{S}_1(\xi)\e^{-\chi_{1,-}(\xi)/\delta} + \mathcal{S}_2(\xi) \e^{-\chi_{1,+}(\xi)/\delta}\right] + \mathrm{c.c.} \quad \mathrm{as} \quad \delta \rightarrow 0,
\end{equation}
where $\chi_{1,-}$ is given by the integral \eqref{3.chiint}, and can be approximated in the long-wave limit using the value of $y_0(\xi)$ given in \eqref{2.y0z0}. The value of $\chi_{1,+}$ can be written as an integral and approximated in near-identical fashion.

Using the remainder equations \eqref{4.truncgov1}--\eqref{4.truncgov2}, we can also show that
\begin{align}\nonumber
y_{\mathrm{exp}}(\xi) \sim \frac{\pi\Lambda \i \delta^{5/2}}{2c_{\epsilon}^2\sqrt{\chi'_{1,-}(\xi+1)}} &\left[ \mathcal{S}_1(\xi+1)\e^{-\chi_{1,-}(\xi+1)/\delta} + \mathcal{S}_2(\xi+1) \e^{-\chi_{1,+}(\xi+1)/\delta}\right]\\
+\frac{\pi\Lambda \i \delta^{5/2}}{2c_{\epsilon}^2\sqrt{\chi'_{1,-}(\xi-1)}} &\left[ \mathcal{S}_1(\xi-1)\e^{-\chi_{1,-}(\xi-1)/\delta} + \mathcal{S}_2(\xi-1) \e^{-\chi_{1,+}(\xi-1)/\delta}\right]+ \mathrm{c.c.} \quad \mathrm{as} \quad \delta \rightarrow 0.\label{4.yexp}
\end{align}

Consequently, we have obtained an asymptotic description for the exponentially small far-field oscillations present in the travelling wave solution to \eqref{1.FPUT} in the small mass ratio limit.

\textcolor{black}{It is very straightforward to obtain the form of symmetric two-sided nanopteron solutions, which are steady rather than simply metastable. In this case, the exponentially small contributions are given by \eqref{4.zexp}--\eqref{4.yexp}, with the change that $\mathcal{S}_j$ switches from $-1/2$ to $1/2$ as the corresponding Stokes lines are crossed, rather than zero to one. This produces symmetric oscillations on both sides of the wavefront, leading to a system in which the wave has both an energy source ahead of the wavefront, and an energy sink in its wake. These are asymptotic representations of true travelling waves that do not radiate energy as they propagate.}
 
\subsection{Far-field oscillations}\label{S.FarField}
 
It is possible to use \eqref{3.chieq} to simplify the form of $z_{\mathrm{exp}}$ in the far field, corresponding to $\xi \rightarrow -\infty$. We note that that $y_0(\xi+1) - y_0(\xi-1)$ decays exponentially in this limit, as the leading-order solution is known to be a true solitary wave. This implies that $\chi' \sim {\i \sqrt{2}}/{c_{\epsilon}}$ as  $\xi \rightarrow -\infty$. 

\begin{figure}
\centering
\begin{tikzpicture}
[xscale=1,>=stealth,yscale=1]

\draw[line width=2pt,black] (-2,1.8) -- (-2,0) node[below] {\scriptsize{$-1$}} -- (3,0);
\draw[line width=2pt,black] (2,1.8) -- (-2,1.8);
\draw[line width=3pt] (-2+0.15,1.8+0.15) -- (-2-0.15,1.8-0.15);
\draw[line width=3pt] (-2-0.15,1.8+0.15) -- (-2+0.15,1.8-0.15);
\draw[line width=3pt] (2+0.15,1.8+0.15) -- (2-0.15,1.8-0.15);
\draw[line width=3pt] (2-0.15,1.8+0.15) -- (2+0.15,1.8-0.15);
\fill[black] (3,0) circle (0.1);

\draw[->] (-4,0) -- (4,0) node[right] {\scriptsize{$\mathrm{Re}(s)$}};
\draw[->] (0,-1) -- (0,2.5) node[above] {\scriptsize{$\mathrm{Im}(s)$}};

\node at (-2,2) [above] {\scriptsize{$\xi_{1,-} = \frac{ \i\pi}{\sqrt{6}\epsilon} - 1$}};
\node at (2,2) [above] {\scriptsize{$\xi_{1,+} = \frac{ \i\pi}{\sqrt{6}\epsilon} + 1$}};
\node at (3,-0.05) [below] {\scriptsize{$\xi$}};
\node at (-2,0.9) [left] {\scriptsize{$\mathcal{C}_1$}};
\node at (1,0) [below] {\scriptsize{$\mathcal{C}_2$}};
\node at (1,1.8) [below] {\scriptsize{$\mathcal{C}_0$}};

\end{tikzpicture}

\caption{The integral contour for \eqref{3.chiint}, which connects the singularity location $s = \xi_{1,+}$ with $s = \xi$. The location of the singularity is denoted by a cross, while the contour is a thick black line. The integral contour is deformed to pass through $s = \xi_{1,-}$. It may be divided into a vertical component $\mathcal{C}_1$ and two horizontal components, $\mathcal{C}_0$ and $\mathcal{C}_2$. The integral contribution along $\mathcal{C}_1$ is real, and the integral contributions along $\mathcal{C}_0$ and $\mathcal{C}_2$ are imaginary; this implies that $\mathrm{Re}(\chi_{1,+})$ is constant for real-valued $\xi$, and identical to the real part of $\mathrm{Re}(\chi_{1,-})$.} 
   \label{F:Contours2}
\end{figure}
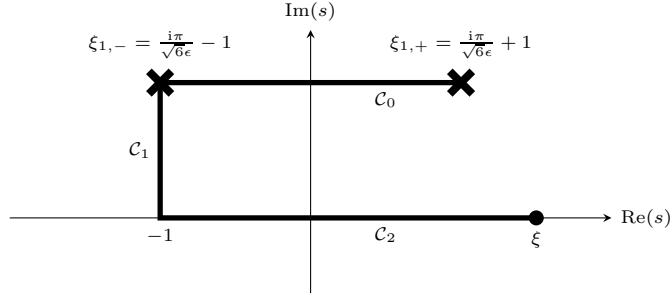

Additionally, we see from contour deformation that $\mathrm{Re}(\chi_{1,-}) = \mathrm{Re}(\chi_{1,+})$, and that this quantity is constant for both singulants. This is determined by deforming the contour for $\chi_{1,+}$ to that depicted in Figure \ref{F:Contours2}. Integrating $\chi'$ along contours $\mathcal{C}_0$ and $\mathcal{C}_2$ produces imaginary contributions, while integrating along $\mathcal{C}_1$ produces the same real contribution as $\chi_{1,-}$. This can be confirmed by comparing the contours in Figure \ref{F:Contours} and Figure \ref{F:Contours2}, and noting that the integrand is identical in both cases.

Using these simplifications, we find that in the far-field where both Stokes multipliers are active, the asymptotic behaviour of the system simplifies to
%\begin{equation}\label{4.zexp2}
%z_{\mathrm{exp}}(\xi) \sim  \frac{\pi\Lambda \sqrt{\i c_{\epsilon}} \delta^{3/2}}{2^{1/4}} \left[\e^{-\chi_{1,-}(\xi)/\delta} +e^{-\chi_{1.+}(\xi)/\delta}\right] + \mathrm{c.c.} \quad \mathrm{as} \quad \delta \rightarrow 0.
%\end{equation}
%We also recall that $\mathrm{Re}(\chi)$ is a positive constant, and identical for both singularities. The far-field behaviour therefore simplifies to give
\begin{equation}\label{4.zexp3}
z_{\mathrm{exp}}(\xi) \sim  \frac{\pi\Lambda \sqrt{\i c_{\epsilon}} \delta^{3/2}}{2^{1/4}} \e^{-\mathrm{Re}(\chi_{1,-})/\delta} \left[ \e^{-\i\,\mathrm{Im}(\chi_{1,-}(\xi))/\delta} + e^{-\i\,\mathrm{Im}(\chi_{1,+}(\xi))/\delta}\right] + \mathrm{c.c.} \quad \mathrm{as} \quad \delta \rightarrow 0.
\end{equation}
Using the result that $\Lambda = 8 \sqrt{\pi\i}/c_{\epsilon}$ and simplifying gives
\begin{equation}\label{4.zexp4}
z_{\mathrm{exp}}(\xi) \sim \frac{16\pi^{3/2}\delta^{3/2}}{c_{\epsilon}^{1/2}2^{1/4}} \e^{-\mathrm{Re}(\chi_{1,-})/\delta} \left[\sin\left(\frac{\mathrm{Im}(\chi_{1,-}(\xi))}{\delta}\right) + \sin\left(\frac{\mathrm{Im}(\chi_{1,+}(\xi))}{\delta}\right)\right] \quad \mathrm{as} \quad \delta \rightarrow 0.
\end{equation}
Recall that $\chi' \sim \i \sqrt{2}/c_{\epsilon}$ as $\xi \rightarrow -\infty$. We therefore write \eqref{4.zexp4} in the convenient form 
\begin{equation}\label{4.zexpfar}
z_{\mathrm{exp}}(\xi) \sim \frac{16\pi^{3/2}\delta^{3/2}}{c_{\epsilon}^{1/2}2^{1/4}} \e^{-\mathrm{Re}(\chi_{1,-})/\delta} \left[\sin\left(\frac{\sqrt{2}\xi}{\delta c_{\epsilon}} + \phi_{1,-}\right) + \sin\left(\frac{\sqrt{2}\xi}{\delta c_{\epsilon}} + \phi_{1,+}\right)\right] \quad \mathrm{as} \quad \delta \rightarrow 0,
\end{equation}
where $\phi_{1,\pm}$ is a constant phase offset. It is straightforward to find a similar expression for $y_{\mathrm{exp}}$ in near-identical fashion.

We see that $z_{\mathrm{exp}}$ contains two non-decaying wavetrains as $\xi \rightarrow -\infty$, far behind the wave front. These wave trains have high frequency and identical exponentially small amplitude in the limit that $\delta \rightarrow 0$. Calculating the phase offsets of the wavetrains requires direct evaluation of the integral \eqref{3.chiint}, and plays an important role in finding the orthogonality condition in Section \ref{S.Orthogonality}. Finding the orthogonality condition amounts to determining mass ratios that produce phase offsets for which the two wavetrains interfere destructively, eliminating the far-field oscillations.

\textcolor{black}{In the case of steady symmetric two-sided nanoptera, $\mathcal{S}_j$ switches from $-1/2$ to $1/2$ as the Stokes lines are crossed. This means that the amplitude of the corresponding waves is half of the far-field amplitude for the metastable one-sided nanopteron, given in \eqref{4.zexpfar}, and that the waves extend symmetrically in both directions.}

\section{Numerical comparisons}\label{S.Numerics}

In order to determine the utility of the asymptotic description of far-field oscillations in \eqref{4.zexp}, we compare the amplitude predicted by the asymptotic analysis with a numerical study of the diatomic FPUT travelling wave. \textcolor{black}{While it may seem sensible to simulate the steady two-sided nanoptera, these waves do not appear in simulations with localized initial data.  It is possible to produce such waves with carefully constructed numerical schemes, as in \cite{Boyd6}, in which symmetric oscillatory basis functions are used to capture the two-sided oscillations.} 

\textcolor{black}{Fortunately the radiation of energy in the metastable one-sided system occurs on a sufficiently slow slow time-scale that the transient effects vanish substantially before the wave decay has any measurable effect on the amplitude of the oscillations. This time scale is conjectured in \cite{Giardetti1} to be exponentially large in the small mass ratio parameter. Consequently, we compute the one-sided metastable nanopteron solution with localized initial data, and determine the amplitude in this ``quasi-steady'' phase, in which transient effects have disappeared, but there is not yet any apparent decay in the amplitude of the oscillations. }

The numerical method was implemented in \textsc{Matlab} using an implementation of the fourth-order Runge Kutta algorithm (RK4). Rather than computing the particle positions directly, the implementation computed the quantities $r(n,t)$ and $p(n,t)$ described in \eqref{1.RP}. The particle positions $y(n,t)$ and $z(n,t)$ were obtained by inverting the relationship between these quantities. This implementation was chosen because the leading-order behaviours $r_0(n,t)$ and $p_0(n,t)$ are both zero in the far field, which is computationally convenient. The initial conditions were chosen to be $r(n,0) = r_0(n,0)$ and $p(n,0) = p_0(n,0)$.

The domain was restricted to include $M = 2^{12}$ particles with indices given by $-M/2+1 \leq n \leq M/2$, with periodic boundary conditions. The initial condition was given by the leading-order travelling wave solution \eqref{2.y0z0}, and the time step was chosen to be $h = 1/40$. In order to prevent interactions between the far-field oscillations and the leading-order travelling wave as it returns to its original position, a window function was applied to the solution. This involved multiplying $r(n,t)$ and $p(n,t)$ by a function $W(n - n_{\mathrm{front}} + M/8)$, where 
\begin{equation}
W(k) = \left\{
        \begin{array}{ll}
            1, & \quad |k| \leq \tfrac{5N}{16}, \\
            1 - \tfrac{8}{N}(|k| - \tfrac{5N}{16}), & \quad \tfrac{5N}{16} < |k| \leq \tfrac{7N}{16}, \\
            0, & \quad\tfrac{7N}{16} < |k| \leq \tfrac{N}{2}.
        \end{array}
    \right.
\end{equation}

\textcolor{black}{It is important to note that this windowing process necessarily causes the computed solution to not conserve energy. A more robust scheme would involve performing simulations on a very large domain, in which the edges are sufficiently far from the wave front that even a disturbance travelling at the speed of sound in the system, $c_0$, would be unable to reach the travelling wave within the simulated timespan. Through direct comparison with such systems, it is possible to conclude that windowing does not measurably affect the amplitude of the exponentially small oscillations.}

\textcolor{black}{The rationale that windowing will not disturb the main wave or its oscillations was given in \cite{Giardetti1}, which cites the study of travelling waves in the FPUT lattice given in \cite{Friesecke2}. This study showed that a localized perturbation to a travelling wave in a monoatomic FPUT lattice generates radiation that travels slower than the wave itself. If one interprets the trailing window edge as an energy sink, this indicates that any effects in the solution caused by this energy sink should travel slower than the wave, and hence the edge of the windowed region. }

\textcolor{black}{The argument from \cite{Friesecke2} applies only to monoatomic lattices. It is possible that this argument be adapted to describe the effect of localized perturbations in diatomic chains; however, it is beyond the scope of this study. Instead, we rely on direct comparison between representative windowed and un-windowed simulations, which showed no discernible difference in amplitude during the quasi-steady phase when windowing was applied to the system.}

%In [CITE - Friesche 2002], the effect of perturbations to a travelling wave in the FPUT lattice were studied. It was found that a localized perturbation to the travelling wave would split into two parts: the first is a small perturbation to the travelling wave speed and phase, and the second is radiation that travels slower than the wave front. This is not precisely the system under consideration in the present study; however, it suggests that disturbances to the system caused by windowing effects will also travel more slowly than the wave front. }

%\textcolor{black}{As noted, the system studied here is not identical to the system discussed in \cite{Friesecke2}, which was monoatomic. Consequently, additional analysis would be required in order to draw rigourous conclusions regarding the effect of disturbances, which is substantially beyond the scope of this study. Instead, representative comparisons between wavefronts computed in a periodic windowed domain and identical wavefronts on a sufficiently large non-periodic domain produced indistinguishable wave profiles over the simulation time. This certainly does not constitute a proof that the numerical method is accurate, but it supports the notion that the method is capable of simulating the metastable wave front.}

This windowing applied in the numerical algorithm causes the system to fail to conserve energy; however, this did not cause apparent problems for the algorithm, which was still able to obtain the exponentially small oscillations after transient effects in the system dissipated. In order to eliminate transient effects, the computations were run for $0 < t < t_{\mathrm{max}}$, where typical values of the maximum time $t_{\mathrm{max}}$ were chosen to be between $10^5$ and $10^6$, depending on the time required for transient effects to vanish. In each case, there was no measurable decay in the oscillation amplitudes for $t \leq t_{\mathrm{max}}$.

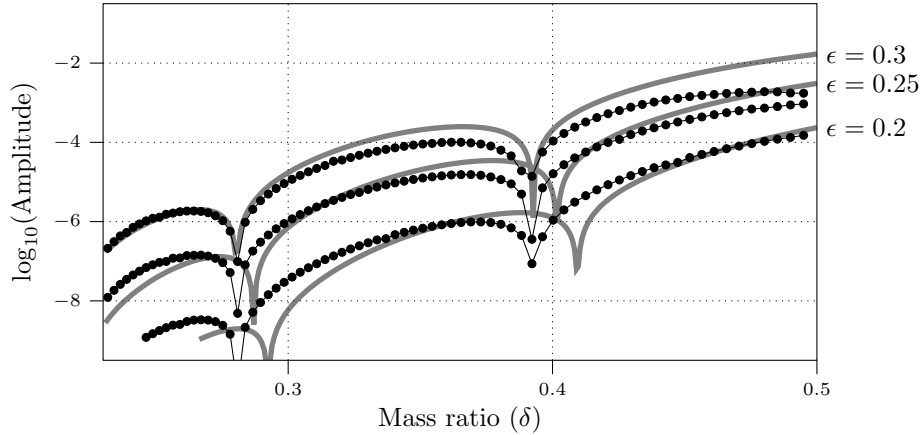
\begin{figure}
\centering
\begin{tikzpicture}
[x=1000*5/5,>=stealth,y=15*5/5]

%\draw[darkgray] plot [smooth] file {EA_k1.txt};
%\draw[darkgray] plot [smooth] file {EA_k0p75.txt};
%\draw[darkgray] plot [smooth] file {EA_k0p5.txt};

\draw[line width=2pt, gray] plot [] file {RNVEC_0p2.txt};
\draw[line width=2pt, gray] plot [] file {RNVEC_0p25.txt};
\draw[line width=2pt, gray] plot [] file {RNVEC_0p3.txt};
%\draw[line width=2pt, gray] plot [] file {RNVEC_0p35.txt};

\draw[] plot [mark=*, mark size=1.5pt] file {AMPVEC_0p2.txt};
\draw[] plot [mark=*, mark size=1.5pt] file {AMPVEC_0p25.txt};
\draw[] plot [mark=*, mark size=1.5pt] file {AMPVEC_0p3.txt};
%\draw[] plot [mark=*, mark size=1.5pt] file {AMPVEC_0p35.txt};

%\draw[] plot [mark=*, mark size=1.5pt] file {RNVEC_0p2.txt};
%\draw[white] plot [mark=*, only marks, mark size=1pt] file {RNVEC_0p2.txt};
%\draw[] plot [mark=*, mark size=1.5pt] file {RNVEC_0p25.txt};
%\draw[white] plot [mark=*, only marks, mark size=1pt] file {RNVEC_0p25.txt};
%\draw[] plot [mark=*, mark size=1.5pt] file {RNVEC_0p3.txt};
%\draw[white] plot [mark=*, only marks, mark size=1pt] file {RNVEC_0p3.txt};

\fill[white] (0.23,-11) -- (0.23,-9.5) -- (0.5,-9.5) -- (0.5,-11) -- cycle;

\draw[dotted] (0.23,-2) -- (0.5,-2);
\draw[dotted] (0.23,-4) -- (0.5,-4);
\draw[dotted] (0.23,-6) -- (0.5,-6);
\draw[dotted] (0.23,-8) -- (0.5,-8);
\draw (0.225,-2) node[left] {\scriptsize{$-2$}} -- (0.23,-2);
\draw (0.225,-4) node[left] {\scriptsize{$-4$}} -- (0.23,-4);
\draw (0.225,-6) node[left] {\scriptsize{$-6$}} -- (0.23,-6);
\draw (0.225,-8) node[left] {\scriptsize{$-8$}} -- (0.23,-8);
%\draw (-2.55,-0) node[left] {\scriptsize{$0$}} -- (-2.5,-0);
%\draw (-2.55,-20) node[left] {\scriptsize{$-20$}} -- (-2.5,-20);
\draw (0.3,-9.5) -- (0.3,-9.85) node[below] {\scriptsize{$0.3$}};
\draw (0.4,-9.5) -- (0.4,-9.85) node[below] {\scriptsize{$0.4$}};
\draw (0.5,-9.5) -- (0.5,-9.85) node[below] {\scriptsize{$0.5$}};

\draw[dotted] (0.3,-9.5) -- (0.3,-0.5);
\draw[dotted] (0.4,-9.5) -- (0.4,-0.5);

%\fill[white] (-2.4,-0.5) -- (-1.6,-0.5) -- (-1.6,-4.5) -- (-2.4,-4.5) -- cycle;
%\draw (-2.4,-0.5) -- (-1.6,-0.5) -- (-1.6,-4) -- (-2.4,-4) -- cycle;
%\draw[darkgray]  (-2.35,-1.5) -- (-2.15,-1.5) node[right] {\scriptsize{Asymptotic}};
%\draw[thick]  (-2.35,-3) -- (-2.15,-3) node[right] {\scriptsize{Numerical}};

\draw node at (0.5,-3.62) [right] {{$\epsilon = 0.2$}};
\draw node at (0.5,-1.77) [right] {{$\epsilon = 0.3$}};
\draw node at (0.5,-2.51) [right] {{$\epsilon = 0.25$}};
%\draw node at (0.5,-1.15) [right] {{$\epsilon = 0.35$}};

\node at (0.365,-11) {{Mass ratio ($\delta$)}};
\node[rotate=90] at (0.2,-5) {{$\log_{10}(\mathrm{Amplitude})$}};

\draw(0.23,-0.5) -- (0.23,-9.5) -- (0.5,-9.5) -- (0.5,-0.5) -- cycle;
\end{tikzpicture}

\caption{Amplitude of the far-field oscillations given by the remainder expression $z_{\mathrm{exp}}$ in \eqref{4.zexp} for a range of values of $\delta$, compared to numerical computations. The thick grey curves represent the amplitude predicted by the leading-order exponentially exponentially small behavior $z_{\mathrm{exp}}$, and the filled circles represent the amplitude obtained using numerical simulations. It is clear that the asymptotic and numerical solutions have the same qualitative behaviour, and are a reasonable quantitative match. For smaller values of $\delta$ than those depicted, it became challenging to isolate the oscillatory behaviour within the computation time window. In the solutions, there clearly exist values of $\delta$ that result in wave cancellation (corresponding to zero values of the amplitude). 
}
   \label{F:Cancellation}
\end{figure}

The results of this numerical computation are illustrated in Figure \ref{F:Cancellation}, in which they are compared to the asymptotic prediction \eqref{4.zexp}. It was numerically challenging to determine wave amplitudes for values of $\delta$ smaller than those presented in Figure \ref{F:Cancellation}. We see that the asymptotic method does capture the qualitative behaviour predicted by the numerical simulations in the region available for comparison. 

While it is challenging to extend the computations to smaller values of $\delta$, it is hoped that smaller values of $\delta$ would correspond to more accurate approximations, as in the case of the diatomic Toda lattice \cite{Lustri5}; however, this could not be shown conclusively without a more accurate numerical study. It can be seen from the larger values of $\delta$ represented in Figure \ref{F:Cancellation} (particular for $\epsilon = 0.25$ and $\epsilon = 0.3$) that the asymptotic approximation becomes qualitatively and quantitatively inaccurate for values of $\delta$ that are not particularly small, as is expected.

There is error introduced into the asymptotic solution by the fact that the leading-order solution \eqref{1.X} is an approximation that depends on a small parameter $\epsilon$, and we expect this to introduce an additional source of error to the asymptotic expansion. Naively, we would expect this error to vary straightforwardly in $\epsilon$, decreasing as $\epsilon \rightarrow 0$; however, from comparing the asymptotic and numeric results, it appears there is exists more complicated nonlinear interaction between the asymptotic error in $\epsilon$ and $\delta$. %This is a potential consequence of the fact that neither $\epsilon$ or $\delta$ is extremely small within the computed numerical range. 

Importantly, we see that both the numerical and asymptotic results predict the existence of mass ratios that cause the leading-order oscillations to cancel entirely, leading to truly localized solitary wave solutions. This corresponds to predictions made in \cite{Lustri5} about the diatomic Toda lattice. We will study these points in more detail in Section \ref{S.Orthogonality}.

\begin{figure}
\centering
\begin{tikzpicture}
[x=125,>=stealth,y=25]
\draw[gray,thick] plot[smooth] file {sineschem.txt}; 
\draw[black] plot[mark=*,only marks,mark size=2pt] file {sineschem2.txt}; 
\draw[->] (0,0) -- (2.5,0) node[right] {\scriptsize{$j$}};
\draw[->] (0,-1.25) -- (0,1.25) node[above] {\scriptsize{$r(j,t_s)$}};
\draw (0.03,1) -- (-0.03,1) node[left] {\scriptsize{$r_{\mathrm{max}}$}};
\draw (0.03,-1) -- (-0.03,-1) node[left] {\scriptsize{$r_{\mathrm{min}}$}};

\draw[black] plot[mark=*,only marks,mark size=1pt] file {numschem_0p14773_e0p3.txt}; 
\draw[->] (0,-3) -- (2.5,-3) node[right] {\scriptsize{$j$}};
\draw[->] (0,-1.25-3) -- (0,1.25-3);
\draw (0.03,-2) -- (-0.03,-2) node[left] {\scriptsize{$1.34\times10^{-4}$}};
\draw (0.03,-4) -- (-0.03,-4) node[left] {\scriptsize{$-1.34\times10^{-4}$}};

\draw[black] plot[mark=*,only marks,mark size=1pt] file {numschem_0p15697_e0p35.txt}; 
\draw[->] (0,-6) -- (2.5,-6) node[right] {\scriptsize{$j$}};
\draw[->] (0,-1.25-6) -- (0,1.25-6);
\draw (0.03,-5) -- (-0.03,-5) node[left] {\scriptsize{$2.28\times10^{-5}$}};
\draw (0.03,-7) -- (-0.03,-7) node[left] {\scriptsize{$-2.28\times10^{-5}$}};

\node at (1.2,-1.5)  {\scriptsize{(a) Schematic: Introduction of periodic effects due to sampling}};
\node at (1.2,-4.5)  {\scriptsize{(b) Simulation: $\epsilon = 0.35$, $\delta = 0.396$}};
\node at (1.2,-7.5)  {\scriptsize{(c) Simulation: $\epsilon = 0.30$, $\delta = 0.384$}};

\end{tikzpicture}

\caption{Schematic (a) illustrates how periodic effects may be appear in regular sampling of fast oscillatory solutions. In this case, the high-frequency continuous gray curve is sampled at regular intervals, indicated by filled black circles. The effect of this sampling is to introduce a slower periodic component into the sampled solution. Figures (b) and (c) show numerically computed wave trains for different values of $\epsilon$ and $\delta$. In (b), there are obvious periodic effects introduced by sampling the high-frequency wave train, but the periodic sampling effects introduced in (c) are less immediately apparent.}\label{F:periodic}
\end{figure}
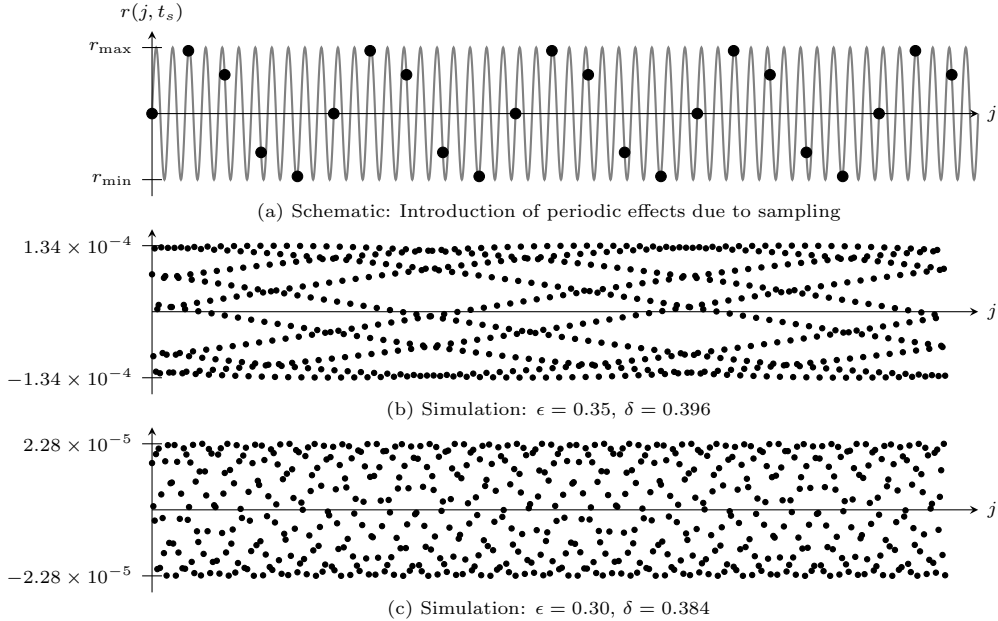

Finally, by looking at individual numerical solutions, we see that sampling the very high frequency sinusoidal term at regular intervals often introduces slower periodic effects into the wave train that are not obvious from the asymptotic form given in \eqref{4.zexp}. This is shown schematically in Figure \ref{F:periodic} (a), representing an example $r(j,t)$ sampled at a particular time. In this schematic, the solid gray curve represents the full solution $r(\xi) = r(j - c_{\epsilon}t)$, where $\xi$ is a continuous variable, while the filled circles represent the sampled points for integer values of $j$ at fixed time $t = t_s$. The fast sinusoidally varying oscillations described in \eqref{4.zexp} are sampled by the discrete chain at regular intervals; as the sampling period is generally not an exact multiple of the sinusoid period, this can introduce slower periodic effects into the solution related to the interaction between the two periodicities.

The numerical wave train in Figure \ref{F:periodic} (b) illustrates visible periodic effects introduced by sampling, while the wave train in Figure \ref{F:periodic} (c) contains more complicated periodic effects. In both cases, it is still possible to calculate the amplitude of the underlying continuous waveform, although care must be taken in order to measure this quantity numerically.  This sampling also explains the relationship between the results obtained in this study and the periodic effects seen in the numerically-computed wave trains in \cite{Giardetti1}. 

%Several examples of exponentially small wave trains in the numerically computed $r(j,t)$ are shown in Figure [REF].
%\textcolor{black}{Should I talk about discrete resonance here, and how I extracted the amplitude? The double period is due to the period of the oscillations v the period of the particles. This is not apparent in the asymptotic analysis, as it is take in the limit that $\delta \rightarrow 0$, and therefore the frequency becomes arbitrarily large.}
%, and if this study was performed with smaller values of $\delta$, it is hoped that the error would decrease as either $\epsilon$ or $\delta$ tended to zero with the other fixed.

\section{Orthogonality Condition}\label{S.Orthogonality}

It is apparent in Figure \ref{F:Cancellation} that there are particular values of $\delta$ which cause the far-field oscillation amplitude to vanish, due to destructive interference between the exponentially small oscillations. As the two wave trains in $z_{\mathrm{exp}}$ given in \eqref{4.zexp} have identical amplitude, it is possible to select parameter values so that the waves precisely cancel each other out. Such configurations produce genuinely localized solitary waves, even at exponentially small orders. Systems with these particular parameter values are said (for example, in \cite{Jayaprakash1,Jayaprakash2}) to satisfy an `anti-resonance' condition. \textcolor{black}{If this condition is satisfied, there are no oscillations generated in the wake of the leading order travelling wave. The wave therefore does not lose energy to an oscillatory wave train, and instead propagates without decaying.}

This behaviour was seen in the diatomic Toda lattice \cite{Lustri5}, and it is possible to show that corresponding behaviour is present in asymptotic solutions the diatomic FPUT lattice. We note that in the wake of the wave front, where all oscillatory contributions are present, the exponentially small contribution is given by
\begin{equation}
z_{\mathrm{exp}} \sim -\frac{\pi\Lambda \i \delta^{3/2}}{\sqrt{\chi'_{1,-}(\xi)}} \left[ \e^{-\chi_{1,-}(\xi)/\delta} + \e^{-\chi_{1,+}(\xi)/\delta}\right] + \mathrm{c.c.} \quad \mathrm{as} \quad \delta \rightarrow 0.
\end{equation}
It is clear that destructive interference occurs if
\begin{equation}\label{5.orth1}
 \e^{-\chi_{1,-}(\xi)/\delta} + \e^{-\chi_{1,+}(\xi)/\delta} = 0.
\end{equation}
We refer to this condition as an `orthogonality condition'. We can simplify this condition by writing the singulant as
\begin{equation}
\chi_{1,+} = \int_{\xi_{1,+}}^{\xi_{1,-}} \chi'(s)\d s + \chi_{1,-} = \lambda + \chi_{1,-}.
\end{equation}
This is a consequence of comparing the contour for $\chi_{1,+}$ shown in Figure \ref{F:Contours2} with the contour for $\chi_{1,-}$ shown in Figure \ref{F:Contours}. The orthogonality condition \eqref{5.orth1} now simplifies to give
\begin{equation}
1 + \e^{-\lambda/\delta} = 0.
\end{equation}
This gives the condition $\lambda/\delta = (2K+1)\pi\i$, for $K \in \mathbb{Z}$, or 
\begin{equation}\label{5.canc-}
\delta = \frac{\sqrt{2}}{(2K+1)\pi c_{\eps}}\int_{\xi_{1,+}}^{\xi_{1,-}} \sqrt{1 + y_0(\xi+1) -y_0(\xi-1)   } \d s,
\end{equation}
which can be approximated in the long-wave limit as
\begin{equation}\label{5.canc}
\delta \approx \frac{1}{(2K+1)\pi}\int_{\pi \i/\sqrt{6} \epsilon - 1}^{\pi \i/\sqrt{6} \epsilon+1 } \sqrt{1 + {\sqrt{6}\,\epsilon} \tanh\left(\sqrt{\tfrac{3}{2}} \epsilon \,(\xi+1) \right) - {\sqrt{6}\,\epsilon} \tanh\left(\sqrt{\tfrac{3}{2}} \epsilon \,(\xi-1) \right)  } \d s,\quad K \in \mathbb{R}.
\end{equation}

While this integral does not have a convenient solution like the corresponding integral in the Toda problem in \cite{Lustri5}, it is straightforward to show that the integrand is real along the integral contour between the two singular points, and consequently that the result of the integral is real-valued. This therefore gives a set of real values for $\delta$ at which the far-field oscillations vanish.

%Recall that $c_{\eps} = \sqrt{2} + \mathcal{O}(\epsilon^2)$, we find that this condition is, to leading order as $\delta \rightarrow 0$, given by
%\begin{equation}
%\delta \sim \frac{1}{(2N+1)\pi }\int_{\pi \i/\sqrt{6} \epsilon - 1}^{\pi \i/\sqrt{6} \epsilon+1 } \sqrt{1 + {\sqrt{6}\,\epsilon} \tanh\left(\sqrt{\tfrac{3}{2}} \epsilon \,(\xi+1) \right) - {\sqrt{6}\,\epsilon} \tanh\left(\sqrt{\tfrac{3}{2}} \epsilon \,(\xi-1) \right)  } \d s,
%\end{equation}
%as $\delta$, $\epsilon \rightarrow 0$. The integral can be asymptotically approximated to two in the limit that $\epsilon \rightarrow 0$; however, this asymptotic approximation is quite slow (as the asymptotic error is proportional to $\epsilon^{1/2}$), and it is more useful to evaluate the integral numerically for purposes of obtaining the orthogonality condition.

In Table \ref{tableref}, asymptotic predictions of the values of $\delta$ that cause the far-field oscillations to cancel are compared with numerical predictions. These numerical predictions were found using the methods from Section \ref{S.Numerics}, however the size of the time steps were decreased in size, and the maximum simulation time was increased in order to resolve the waves at  smaller values of $\delta$ than those presented in Figure \ref{F:Cancellation}. This allowed us to determine the values of $\delta$ corresponding to $K = 2,\ldots,6$.

It is clear from Table \ref{tableref} that the asymptotic predictions of nanopteron-free values of $\delta$ may be accurately predicted using the asymptotic form of $z_{\mathrm{exp}}$. \textcolor{black}{These values correspond to localized solitary waves that propagate without decaying due to energy radiating into the exponentially small oscillations in the wake of the wave front. Furthermore, we note that the analysis which produced these values does not depend on whether the original nanopteron solution under consideration is one-sided (and hence only metastable) or two-sided; both wave families produce truly localized waves for these parameter values.}

  \begin{table}
\centering
  \begin{tabular}{ | c || c | c | }
    \hline
    $K$ & Computed $\delta$  & Asymptotic $\delta$ \\ \hline\hline
    2 & 0.3922 & 0.4258 \\ \hline
    3 & 0.2809 & 0.3041 \\ \hline
    4 & 0.2248 & 0.2365   \\ \hline
    5 & 0.1932 & 0.1935   \\ \hline
    6 & 0.1677 & 0.1638  \\ \hline
  \end{tabular}

\caption{Comparison between $\delta$ values for nanopteron-free solutions computed with $\epsilon = 0.25$, and the asymptotic prediction obtained from \eqref{5.canc} for corresponding values of $K$. The asymptotic prediction provides a good approximation for the numerically-obtained solutions. As in the amplitude predictions, this is not as accurate as the predictions for the diatomic Toda lattice from \cite{Lustri5}, even for larger values of $K$ (and hence smaller values of $\delta$).
}\label{tableref}
\end{table}

\section{Discussion and Conclusions}\label{S.Conclusions}

In this study, we considered travelling slightly supersonic wave behaviour in a diatomic FPUT lattice with small mass ratio. The asymptotic solutions of this system are nanoptera, or solitary-wave solutions that are not exponentially localized, but instead possess trains of oscillations in the far field behind the wave front. These oscillations are exponentially small, so their dynamics are invisible to ordinary asymptotic power-series approaches. The existence of these oscillations was previously proven in \cite{Faver2}.

We demonstrated that far-field oscillations `switch' across special curves in the complex plane known as Stokes curves, which originate at singularities in the analytic continuation of the leading-order behaviour. The far field oscillations present in the asymptotic wave behaviour are therefore a consequence of Stokes Phenomenon. We derived asymptotic forms for these exponentially small oscillations, given in \eqref{4.zexp} and \eqref{4.yexp}, as well as a simplified expression for the behaviour of the oscillations far from the wave front \ref{4.zexpfar}. We compared the results of this analysis to numerical studies, and found that the asymptotic results were useful for predicting both the amplitude of the oscillations as well as the special mass ratios that produce localized solitary wave solutions.

One significant difference between the travelling wave solutions in the diatomic FPUT lattice as opposed to the previous analysis of the diatomic Toda lattice is that the leading order solution could not be determined exactly for the diatomic FPUT approximation, but was instead approximated using the long-wave limit approximation from \cite{Gaison1}. This change introduced a second small parameter into the problem; however, with careful treatment, it was still possible to determine the leading-order behaviour of the late-order terms. This work establishes the effectiveness of exponential asymptotics for determining asymptotic nanopteron solutions in other lattice systems for which the leading-order behaviour must be approximated, such as the diatomic Hertizan lattice \cite{Porter1}, the woodpile lattice \cite{Kim1}, or lattices of resonant granular crystals \cite{Vorotnikov1}.

By isolating the exponentially small terms, we found an asymptotic condition that predicted the wave trains would cancel entirely, given in \eqref{5.canc}. If this condition is satisfied, the solutions do not possess a wave train in the wake of the travelling wave front; instead, they consist of a localized solitary wave. Similar results were found using an exponential asymptotic analysis on the diatomic Toda lattice in \cite{Lustri5}, as well as other studies in which special choices of mass ratios and wave parameters produce localized solitary wave solutions \cite{Jayaprakash1, Kevrekidis1, Xu1}. Importantly, as in the diatomic Toda lattice from \cite{Lustri5}, this orthogonality condition arises as a consequence of Stokes Phenomenon, and the precise cancellation of exponentially small wave trains that appear as two different Stokes curves are crossed.

It is important to emphasise that the solution that we discussed in this paper is a formal solution asymptotic solution, and we did not supply a rigorous proof of these results. Faver \& Wright \cite{Faver1} used rigorous estimates for the local leading-order solitary wave to prove the existence of the exponentially small oscillations in a nanopteron solution of FPUT systems. Typically, exponential asymptotic arguments can be made rigorous using the method of Borel transforms in order to replace the divergent tail of the asymptotic series with a quantity that can be bounded rigorously. This is beyond the scope of the present work; however, examples of rigorous exponential asymptotic bounding may be found in, for example, \cite{Bennett1,Berry5,Berry2,Costin3} and others.

\textcolor{black}{This paper considered one-sided nanopteron solutions in detail, while outlining the corresponding results for symmetric two-sided nanoptera. As noted, only the latter is an example of a true solitary wave. The one-sided solution is metastable, as it slowly loses energy to the far field oscillations which causes the wave to eventually decay on a long timescale. This decay is not visible in the computed asymptotic solution describing the leading-order wave or the small oscillations; a valuable direction of future study would be to determine the asymptotic order on which this decay becomes apparent. It would be of interest to determine whether the unsteady behaviour appears directly at subsequent orders of the formal asymptotic wave behaviour, or whether it requires careful application of multiple scale techniques in order to be captured asymptotically.}

\section{Acknowledgements}

CJL thanks Prof. J. Douglas Wright for helpful discussions on the implementation of the numerical methods applied in Section \ref{S.Numerics}, and Dr Justin Tzou for discussions about the manuscript. CJL is supported by ARC Discovery Project DP190101190.

\bibliography{sydrefs2.bib}
\bibliographystyle{amsplain}

\appendix

\section{Determining $\Lambda$}\label{A.Lambda}

To determine the value of $\Lambda$, we match the late-order expansion in the outer region with the local solution in an inner region near the singularity. Using Van Dyke's matching principle we match the inner limit as $\xi \rightarrow \xi_s$ of the outer expansion with the outer limit of the inner expansion, determined below.

In the inner region near the singularity at $\xi = \xi_s$, we find that
\begin{align}
y_0(\xi +1) &\sim \frac{2}{\xi-\xi_s} + \mathcal{O}(\xi-\xi_s)\label{A.y0p1},\\
y_0(\xi - 1) &\sim -\sqrt{6} \epsilon\, \mathrm{coth}(\sqrt{6}\epsilon)+ \mathcal{O}(\xi-\xi_s)\label{A.y0m1},\\
z_0(\xi) & \sim \frac{1}{\xi-\xi_s} -\frac{\sqrt{6} \epsilon}{2} \mathrm{coth}(\sqrt{6}\epsilon) + \mathcal{O}(\xi-\xi_s).
\end{align}

In order to locate the inner region, we must determine the region in which the inner analysis breaks down. From the form of the late-order ansatz, we find that this occurs at $\delta^2\chi^{-2} = \mathcal{O}(1)$ as $\delta \rightarrow 0$, or $\delta^2 (\xi-\xi_s)^{-1} = \mathcal{O}(1)$. This corresponds to the inner scaling $\xi - \xi_s = \delta^2 \overline{\xi}$. The appropriate rescaled inner variables are given by
\begin{align}
y(\xi+1) = \frac{2}{\delta^2 \overline{\xi}} + \hat{y}(\overline{\xi}+\delta^{-2}),\qquad y(\xi - 1) = \hat{y}(\overline{\xi}-\delta^{-2}),\qquad z(\xi) = \frac{1}{\delta^2  \overline{\xi}} +  \frac{\hat{z}(\overline{\xi})}{\delta^2}.
\end{align}
Retaining the leading-order terms as $\delta \rightarrow 0$ in the rescaled inner equation gives
\begin{equation}
\frac{2}{ \overline{\xi}^3} + \diff{^2 \hat{z}(\overline{\xi})}{\overline{\xi}^2} = -\frac{4 \hat{z}(\overline{\xi})}{c_{\eps}^2 \overline{\xi}}.
\end{equation}
We express $\hat{z}$ in terms of the local series
\begin{equation}
\hat{z}(\overline{\xi}) \sim \sum_{j=1}^{\infty} \frac{a_j}{\overline{\xi}^{j+1}}\quad \mathrm{as} \quad \overline{\xi} \rightarrow 0,
\end{equation}
as the term with power $j=1$ is already built into the inner form of $v(\xi)$. This gives
\begin{equation}
\frac{2}{ \overline{\xi}^3} + \sum_{j=1}^{\infty} \frac{(j+1) (j+2) a_j}{\overline{\xi}^{j+3}} = -\frac{4}{c_{\eps}^2}\sum_{j=1}^{\infty} \frac{a_j}{\overline{\xi}^{j+2}}.
\end{equation}
From matching at leading order, we see that $a_1 = -c_{\eps}^2/2$. At subsequent orders, we obtain the recurrence relation
\begin{equation}
c_{\eps}^2 j (j+1) a_{j-1} = -4a_j.
\end{equation}
Noting the form of $a_1$, this gives
\begin{equation}
a_j = -\frac{c_{\eps}^2}{2} \left(\frac{c_{\eps}^2}{4}\right)^{j}(-1)^j\Gamma(j+2)\Gamma(j+1).
\end{equation}
By comparing the series expression with the inner limit of the late-order ansatz, we find using Stirling's formula 
\begin{align}
\Lambda &= \frac{2^{1/2}}{c_{\eps}^{1/2}} \lim_{r \rightarrow \infty} \frac{a_r (4\i/c_{\eps})^{2r + 5/2}}{\Gamma(2r + 5/2)}= \frac{2^{9/2}\i^{1/2}}{ c_{\eps}} \lim_{r \rightarrow \infty} \frac{ 4^{r}\Gamma(r+2)\Gamma(r+1)}{\Gamma(2r + 5/2)}= \frac{8\sqrt{\pi\i}}{ c_{\epsilon}}.
\end{align}
%We have now completely described the late-order terms of the asymptotic expansion, given by
%\begin{equation}
%v_r(\xi) \sim \frac{8\sqrt{\pi \i}\,\Gamma(2r + 5/2)}{  c_{\eps}\sqrt{\chi'(\xi)}\chi^{2r+5/2}}\quad \mathrm{as} \quad r \rightarrow \infty.
%\end{equation}

\end{document}